\documentclass[a4paper]{article}

\usepackage[affil-it]{authblk}

\usepackage[margin=2.5cm]{geometry}

\usepackage{titlesec}
\usepackage[english]{babel}
\usepackage[utf8]{inputenc}
\usepackage{amsmath}
\usepackage{graphicx}
\usepackage[colorinlistoftodos]{todonotes}
\usepackage{lscape}
\usepackage{multirow}
\usepackage{natbib}

\usepackage{tikz}

\usepackage{subfig}
\usepackage{algorithmicx}
\usepackage{algpseudocode}
\usepackage{algorithm}

\begin{document}

\title{Robust NFP generation for Nesting problems\thanks{This research was partially supported by Project TEC4Growth - RL iMAN - Intelligence for Advanced Manufacturing Systems (NORTE-01-0145-FEDER-000020)}}

\author{Pedro Rocha
	\thanks{Electronic address: \texttt{pedro.f.rocha@inesctec.pt; pfr@isep.ipp.pt}}
    }

\affil{INESCTEC - Instituto de Engenharia de Sistemas e Computadores Tecnologia e Ciência, ISEP - Instituto Superior de Engenharia do Porto, Porto, Portugal}

\date{\today}

\maketitle

\begin{abstract}
Cutting and packing problems arise in a large variety of industrial applications, where there is a need to cut pieces from a large object, or placing them inside a containers, without overlap. When the pieces or the containers have irregular outline, the problem is classified as a Nesting problem. The geometrical challenges of the Nesting problem are addressed by focusing on the geometric aspect of the 2D pieces and containers involved. The challenges of the geometrical component are mainly derived from the complexity of the pieces, due to high number of vertices, which is common when dealing with real world scenarios. This complexity is challenging for current algorithms to process efficiently and effectively, leading to high computational cost and less satisfactory results, particularly when dealing with overlap verification operations. Usually, when tackling Nesting problems, the overlap verification process between two objects is done through the use of a structure known as No-Fit-Polygon (NFP). 

In this work, the generation of the NFP is achieved through a simple algorithm which produces a simplified shape while reducing numerical precision errors and fully representing the region that forms the NFP including positions with perfect fits.

\end{abstract}

\section{Introduction}
\label{sec:intro}

Many problems that arise in several real world scenarios require the placement of pieces inside a container (or cutting pieces from a large object) without overlap and while aiming to minimize wasted space by achieving a compact configuration. If one of these problems deals with irregular pieces or containers it is classified as Nesting problem, which is also known as an Irregular Piece Packing problem. This problem requires an adequate geometric representation of its objects with great accuracy, which usually generates a significant amount of geometric information (vertices, pixels, etc), creating difficulties for current algorithms. One of the main challenges in the geometrical component of the Nesting problem is ensuring the non-overlap and correct containment of the pieces, while maintaining an acceptable level of computational efficiency.  

The overlap verification in Nesting problems is usually achieved by using a geometrical structure known as No-Fit-Polygon (NFP). This structure is a set of feasible placement locations of one polygon relative to another, that allows specifying the regions and placement locations where pieces do not overlap. In this paper, an algorithm is presented that generates the NFP with a reduced number of geometrical components that define it, while using a simplified process that deals with numerical precision problems. This improves the quality of the NFP while also reducing the overhead cost of overlap verification.

This paper is organized in several sections, being this one the introduction. The remainder of this section contains the motivation for this work and main contributions. The following section, the second, present a literature review, making reference to some approaches used generate the NFP focusing on Nesting requirements. The third section contains a description of our approach, while the fourth presents alternative applications for the proposed algorithm followed by fifth section with the conclusions and future work.

\subsection{Motivation}

The proposed approach addresses the challenges of NFP generation, dealing with numerical precision problems and generating a NFP with a reduced amount of geometric components while maintaining its representation quality. Since the NFP is extensively used in approaches addressing Nesting problems, improving it leads to computational efficiency gains and possible improvements in solution quality in cases where a complete NFP representation was not previously used.

Improving the efficiency of the geometrical component leads to more resources being devoted to addressing the combinatorial component of the Nesting problem, which can enable the use of new approaches, and reaching higher quality solutions.

The improvement of current geometric overlap verification methods has a relevant impact on many scientific fields where similar problems arise, considering both economic and environmental aspects.

\subsection{Contribution}

In many real world scenarios, such as industrial applications that deal with leather, furniture, metallurgy, shipbuilding, sheet metal cutting, plastics, and others, arise problems that require placing a set of pieces into a container, in the most efficient non-overlapping configuration that minimizes wasted space. These problems require dealing with geometrical aspects derived from the structure of the pieces and their relative position inside the container, which must dealt with effectiveness and efficiently, while managing the trade off between solution quality and computational cost.

This work has a relevant impact in several industries and scientific fields, at various levels, wherever similar problems arise, producing benefits of economic (such as cost reduction) and environmental (less waste, energy and raw material consumption) importance. It provides an algorithm that is simple to implement, and generates overlap verification structures based on No-Fit-Polygons that are very robust. This has a significant impact by improving current approaches and opening new research paths in previously underdeveloped areas such as 3D irregular packing. 

\section{Literature Review} 
\label{sec:litrev}

The placement of irregular items in the most efficient placement positions inside a container is heavily dependent on the geometrical structures that are used. The items cannot overlap and must completely fit inside the container. In the literature one can find approaches that are based on distinct geometric representations such as grids, circles, direct trigonometry, phi-functions and No-Fit-Polygons (NFP). In \citep{kn:Bennell2008a} can be found an introduction to these different representations focusing specifically on Nesting problems.

The NFP has been a commonly used approach due to its increased efficiency compared to direct trigonometry, having the benefit of being built directly from the edges of the polygons that represent the real objects (\citep{kn:Bennell2008a}). 

\subsection{The No-Fit-Polygon}

\begin{figure}[!htbp]
\centering
	\subfloat[First combination][$NFP_{AB}$\label{subfig:fig26}]{
   		\includegraphics[scale=0.20]{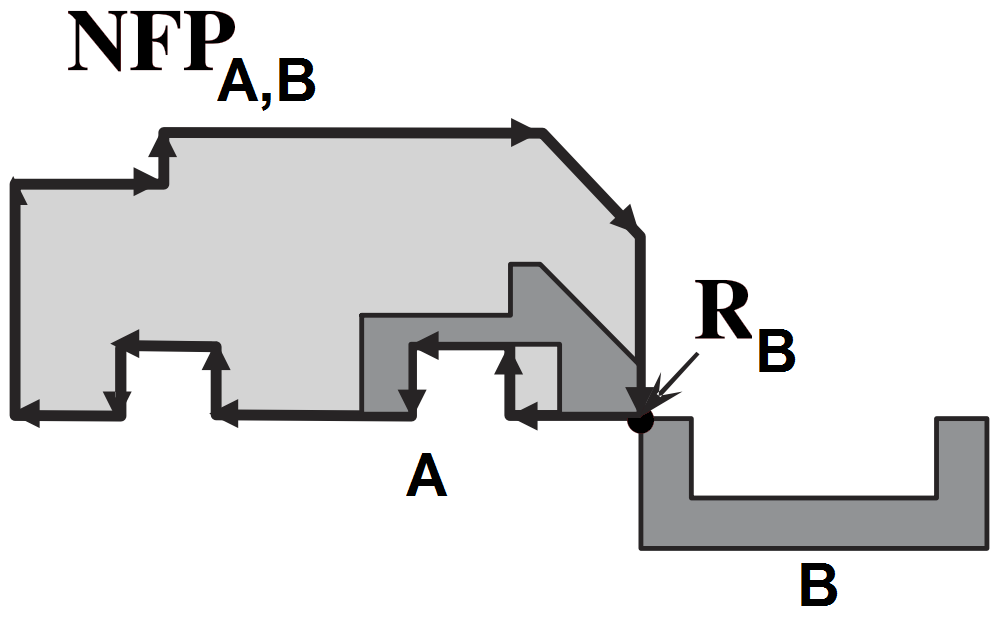} 
 	}
 	\subfloat[Second combination][$IFP_{AB}$\label{subfig:fig27}]{
   		\includegraphics[scale=0.20]{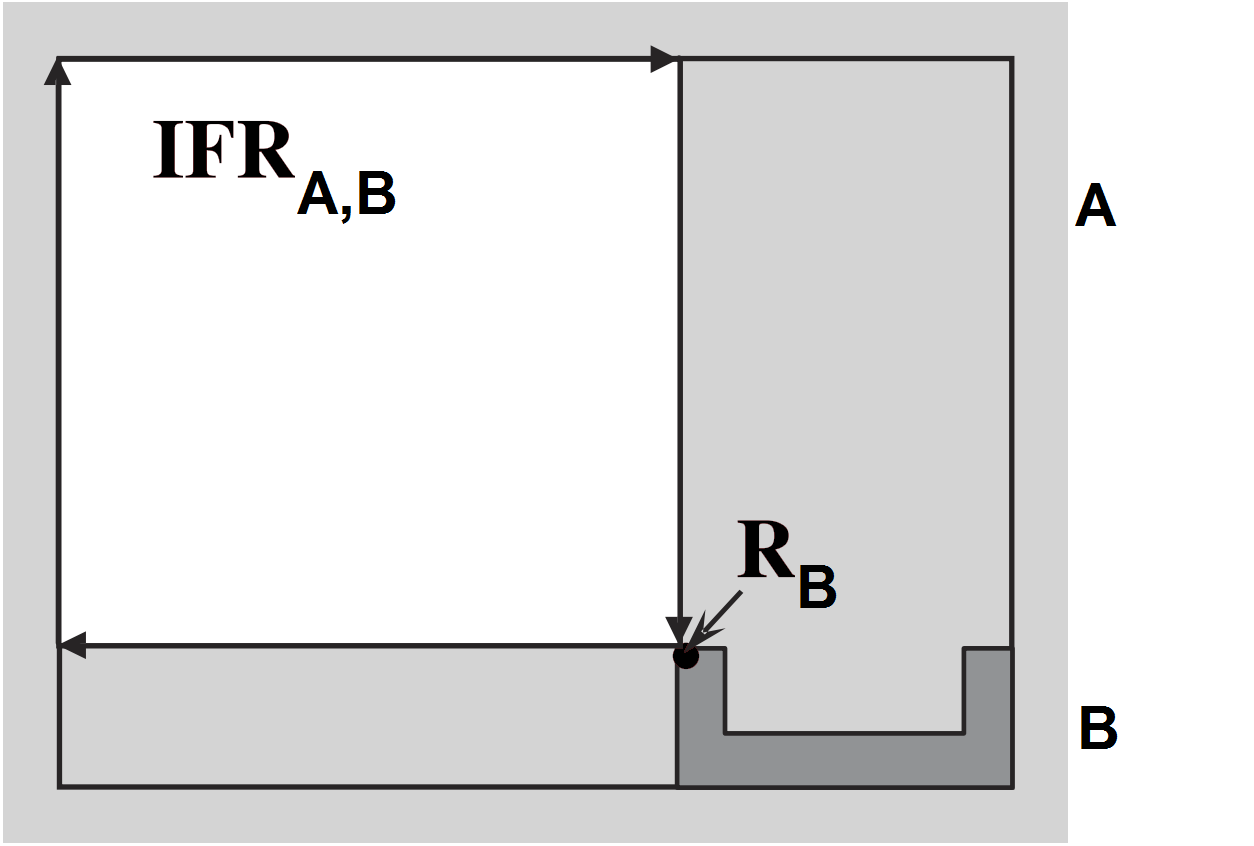}   
 	}
\caption{\label{fig:NFP_IFP}NFP between polygon A and polygon B (a) and IFP between board A and polygon B (b) (adapted from (\citep{kn:Gomes2006}))}
\end{figure}

\begin{figure}[!htbp]
\centering
 \subfloat[First combination][First example\label{subfig:fig28a}]{
   \includegraphics[scale=0.175]{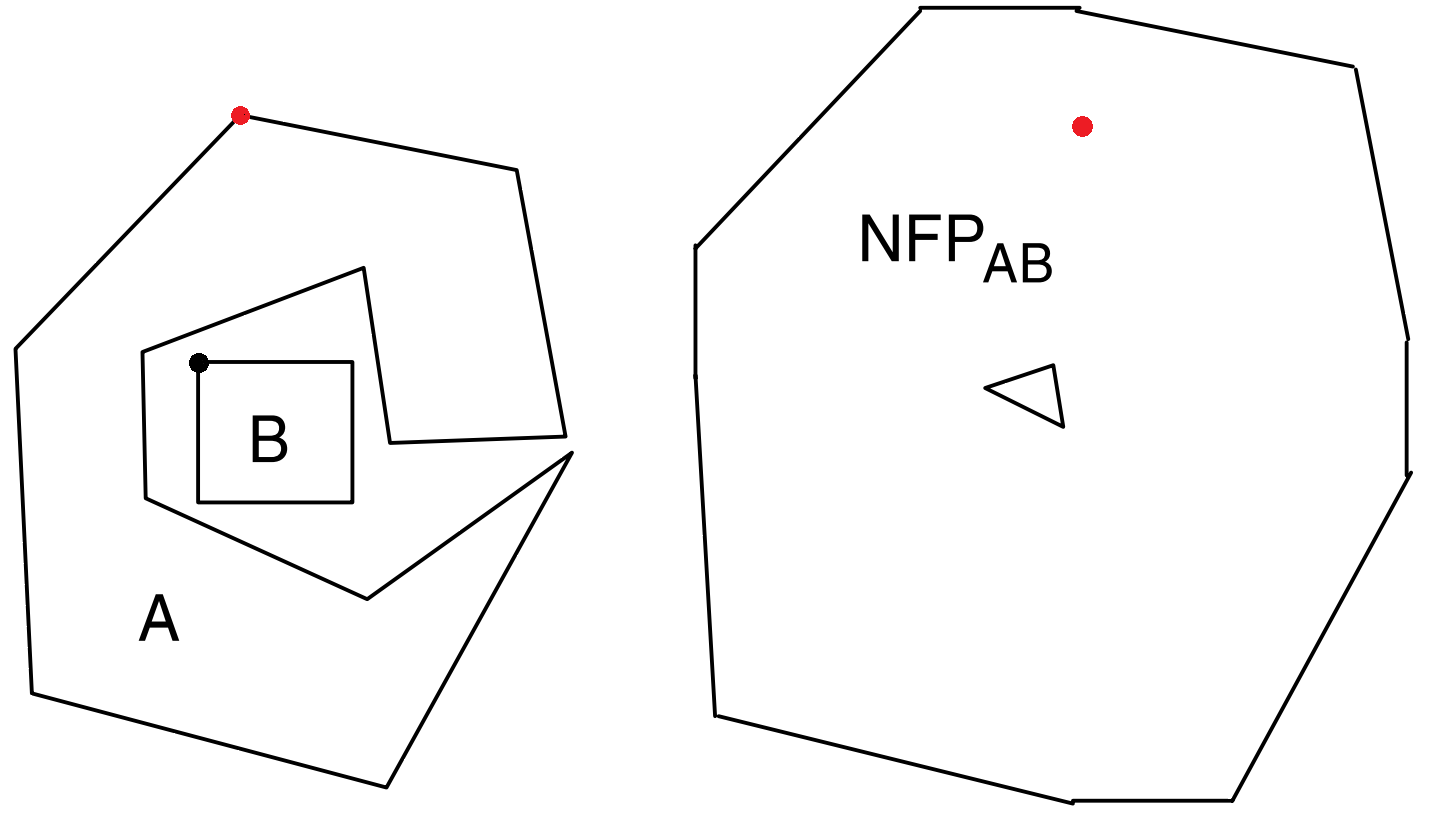} 
 }
 \subfloat[Second combination][Second example\label{subfig:fig28b}]{
   \includegraphics[scale=0.175]{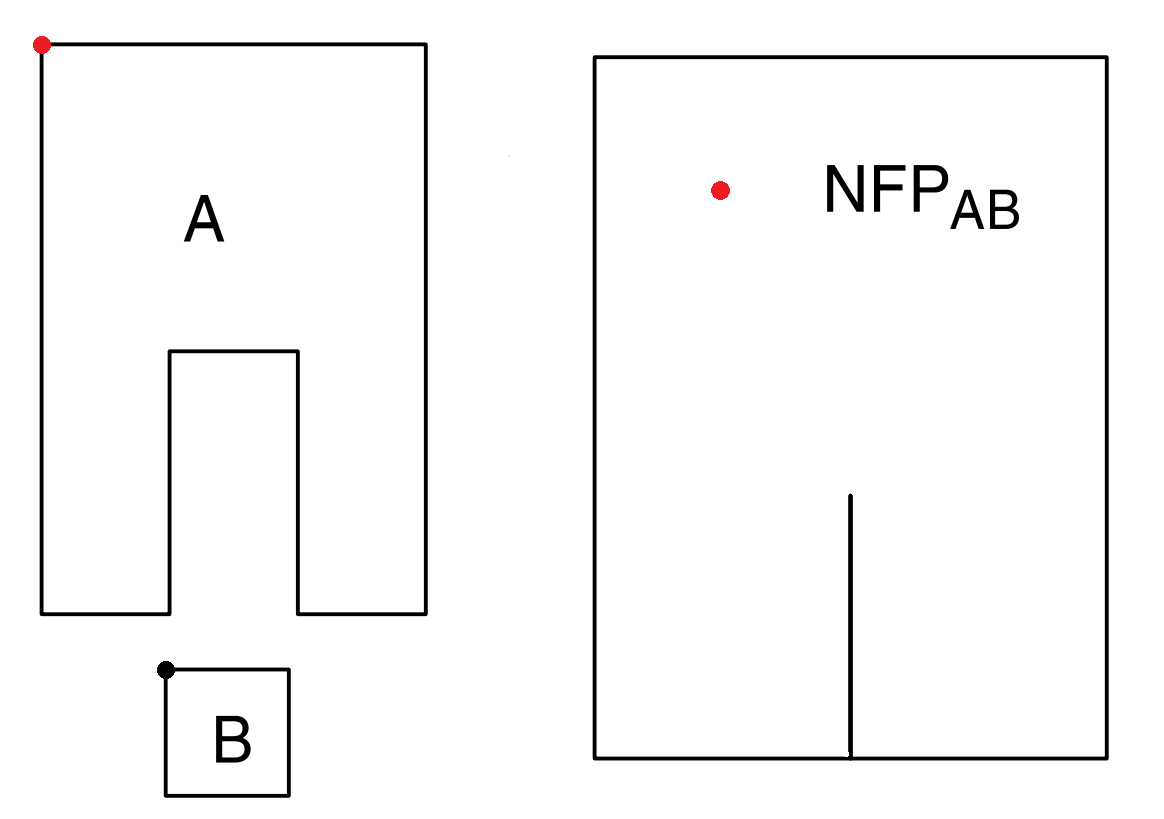}   
 }
 
  \subfloat[Third combination][Third example\label{subfig:fig28c}]{
   \includegraphics[scale=0.20]{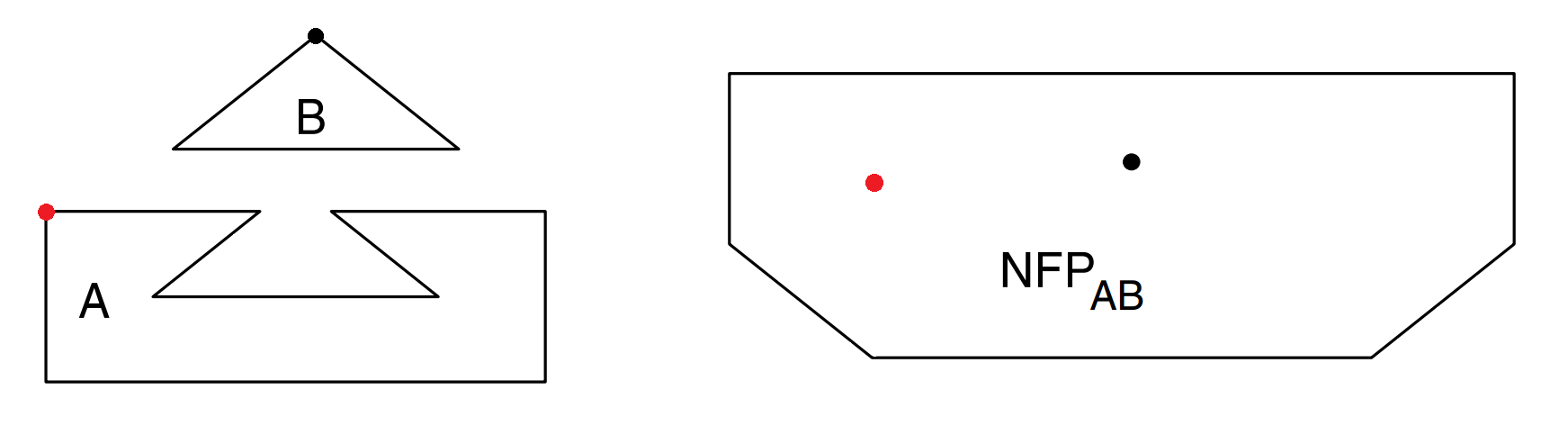}  
 }
\caption[NFPs and degenerate cases]{\label{fig:NFP_dgn}Examples of combinations of polygons that generate NFPs and their degenerated cases (adapted from~\citep{kn:Bennell2009})}
    
\end{figure}

The NFP is a collection of all feasible placement locations of a polygon relative to another. It transforms the overlap verification process between two polygons into the overlap verification between a polygon and a vertex, which is more computationally efficient. Another benefit is that the NFP can be generated at a pre-processing phase, but only for a reduced set of possible piece orientations. In situations where continuous rotations are advantageous, this approach is not adequate due to its significant computational cost generating the NFP for every new orientation of the pieces. One drawback of this approach is also the difficulty in developing a robust approach to generating the NFP from pairs of generic irregular shapes, due to emergence of degenerate cases. When the NFP is generated between a piece and the container, or hole, it is known as the Inner-Fit-Polygon (IFP). An example of a NFP and IFP generated by a sliding algorithm can be seen in Fig.~\ref{fig:NFP_IFP} and an example of its degenerate cases can be seen in Fig.~\ref{fig:NFP_dgn}.

\subsection{No-Fit-Polygon Generation}

In the literature, there are three main approaches dedicated to generate the NFP. They are based on a Sliding Algorithm, Slope Diagrams or Minkowsky Sum and Polygonal Decomposition. There is also another approach that specifies the inverse region of the NFP, defined as Collision Free Region (CFR).

\begin{figure}[!htbp]
\centering
\includegraphics[scale=0.225]{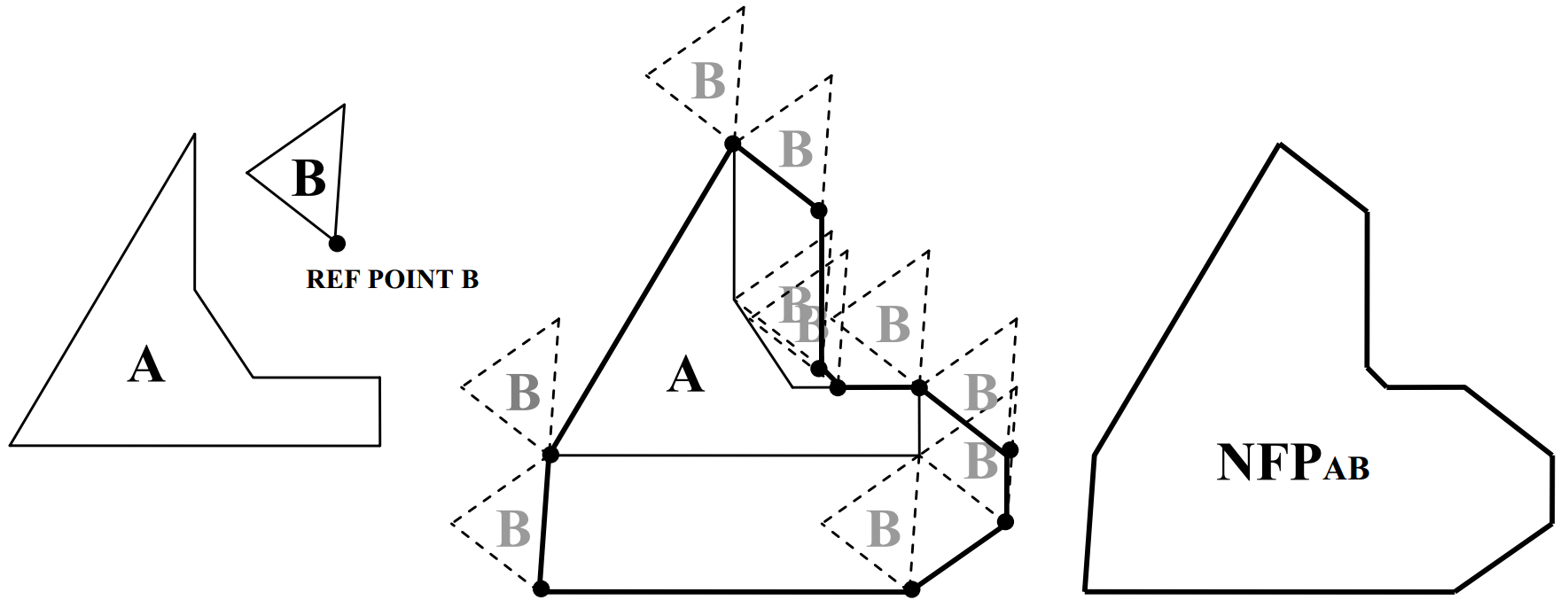}
\caption{\label{fig:NFP_Sliding}Sliding example to create the NFP between piece $A$ and $B$.(\citep{kn:Burke2007}).}
\end{figure}

The sliding algorithm was initially proposed by \citep{kn:Mahadevan1984} and computes the NFP using an approach that simulates the sliding of an orbiting piece around a stationary piece, without ever loosing contact. The NFP is defined by the path created by the reference point of the orbital piece while it slides around the stationary piece until it reaches the initial position. This approach has difficulties dealing with pieces with concavities that have very narrow entries or internal holes. It also has several challenges when the scale of the pieces very large, since the orbital piece may lose contact with the edges of the stationary due to numerical precision errors. Another challenge is the identification of possible placement positions inside holes of the stationary piece, including perfect fit situations. This approach was improved by \citep{kn:Whitwell2004} which tries to address perfect fit sliding situations and an extension by \citep{kn:Burke2006} addresses the identification of holes that have potential for feasible placement positions. This is also explored further in \citep{kn:Burke2007}. The process for generating the NFP using the orbital algorithm can be seen in Fig.~\ref{fig:NFP_Sliding}.

\begin{figure}[!htbp]
\centering
\includegraphics[scale=0.30]{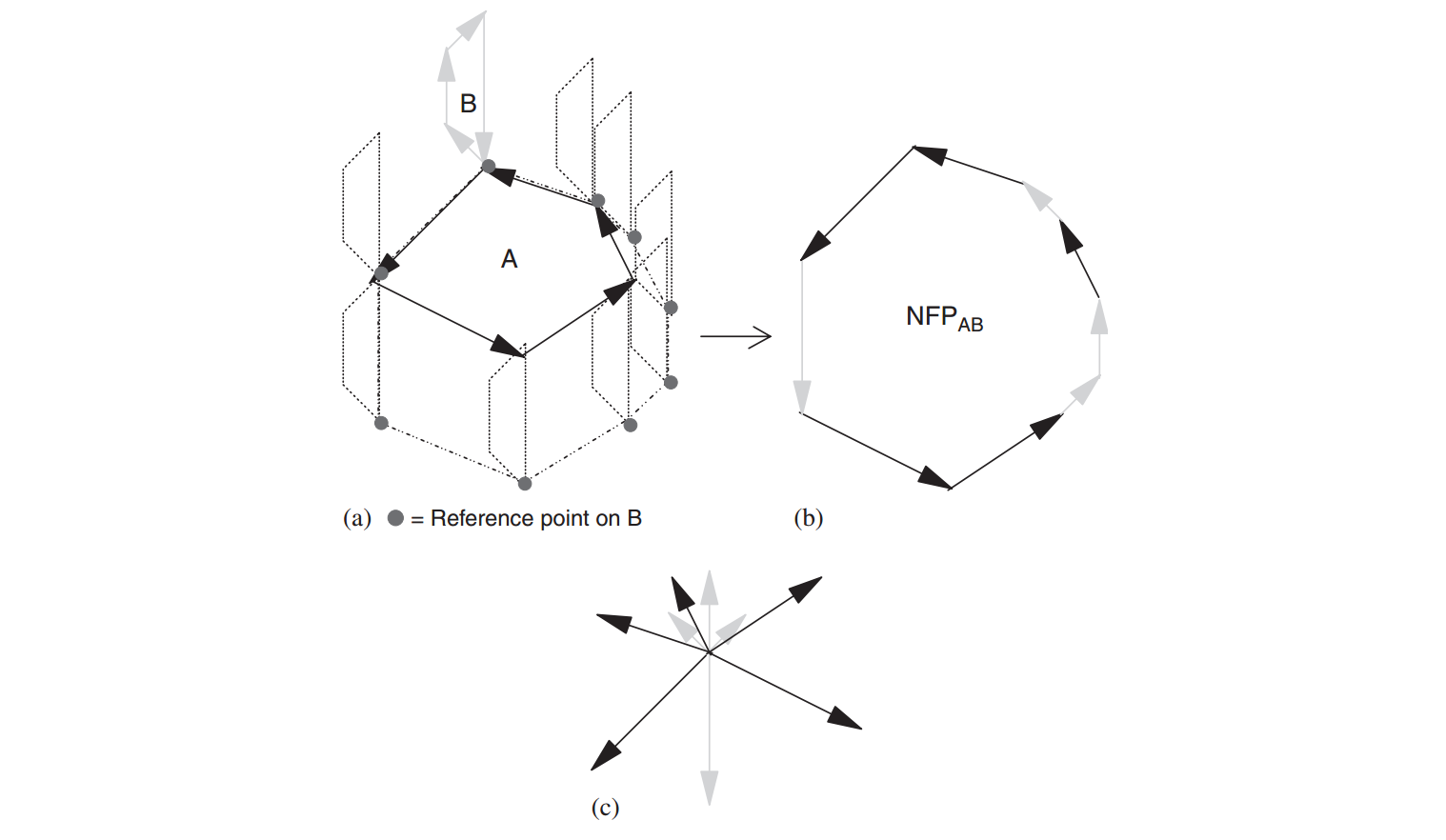}
\caption{\label{fig:NFP_Slope}Slope example to create NFP between piece $A$ and $B$.(\citep{kn:Bennell2008b})}
\end{figure}

The approach based on slope diagrams was proposed by \citep{kn:Cuninghame1989} which computes the NFP using pairs of convex polygons, by using an ordered list of the edges of both polygons ordered by slope. The edges are added sequentially building the NFP until all edges are used. This approach has limited application due to only supporting convex polygons, which was addressed by \citep{kn:Ghosh1991} which extended its use for non-convex cases. When dealing with concavities, the approach does not use a sequence of ordered edges by slope, but it repeats the edges of the convex polygon every time it interacts with a concavity of the other polygon. The limitation of this method appears when the concavities interact with each other. The challenge of two concave polygons was addressed by \citep{kn:Bennell2001} by replacing the concave edges for artificial edges of the orbital polygon, which simulates a convex polygon. The artificial edges are then replaced by the original concave edges including other transversed edges by the stationary polygon. Other improvements can be seen in \citep{kn:Dean2006}, \citep{kn:Bennell2008b}. This approach is also used to generate a concept similar to the NFP known as the Minkowsky sum, as seen in \citep{kn:Milenkovic1991} and \citep{kn:Bennell2001}. An example of the process for generating the NFP using Minkowsky sum can be seen in Fig.~\ref{fig:NFP_Slope}.

\begin{figure}[!htbp]
\centering
\includegraphics[scale=0.325]{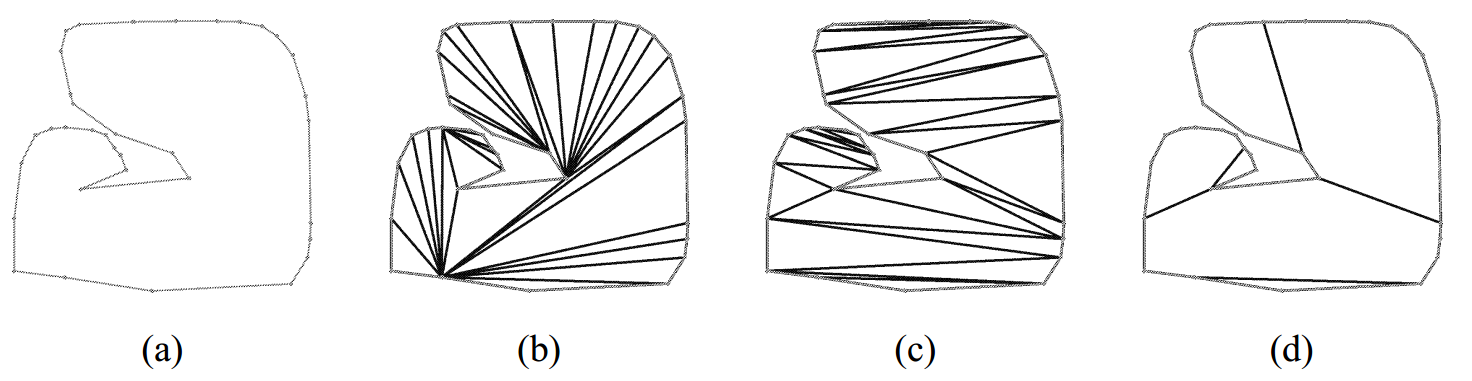}
\caption{\label{fig:NFP_PolDecomp}Decomposition example.(\citep{kn:Agarwal2002})}
\end{figure}

The approach based on polygonal decomposition computes the NFP by decomposing each irregular polygon into convex sub-polygons and then generating the NFP from each pair of convex sub-polygons from different pieces. The sets of convex NFPs for each pair of pieces can be used directly, or they may be merged together to obtain the full NFP from the original polygons. This merging process is the most challenging since its very complicated to merge the polygons while dealing with numerical precision errors and detecting perfect fit or perfect sliding placement positions. One of the challenges is how to produce the minimum number of convex sub-polygons for each polygon, which impacts the total number of NFPs between each pair of pieces, and also their complexity due to the number of vertices of each sub-polygon, depending on the approach used for decomposition. An example with several degrees of polygonal decomposition can be seen in Fig.~\ref{fig:NFP_PolDecomp}. Some authors that use convex decomposition are \citep{kn:Watson1999}, \citep{kn:Babu2001} and \citep{kn:Agarwal2002}. Another similar approach is the decomposition into star-shaped polygons which can be seen in \citep{kn:Li1995}. 

\begin{figure}[!htbp]
\centering
\includegraphics[scale=0.20]{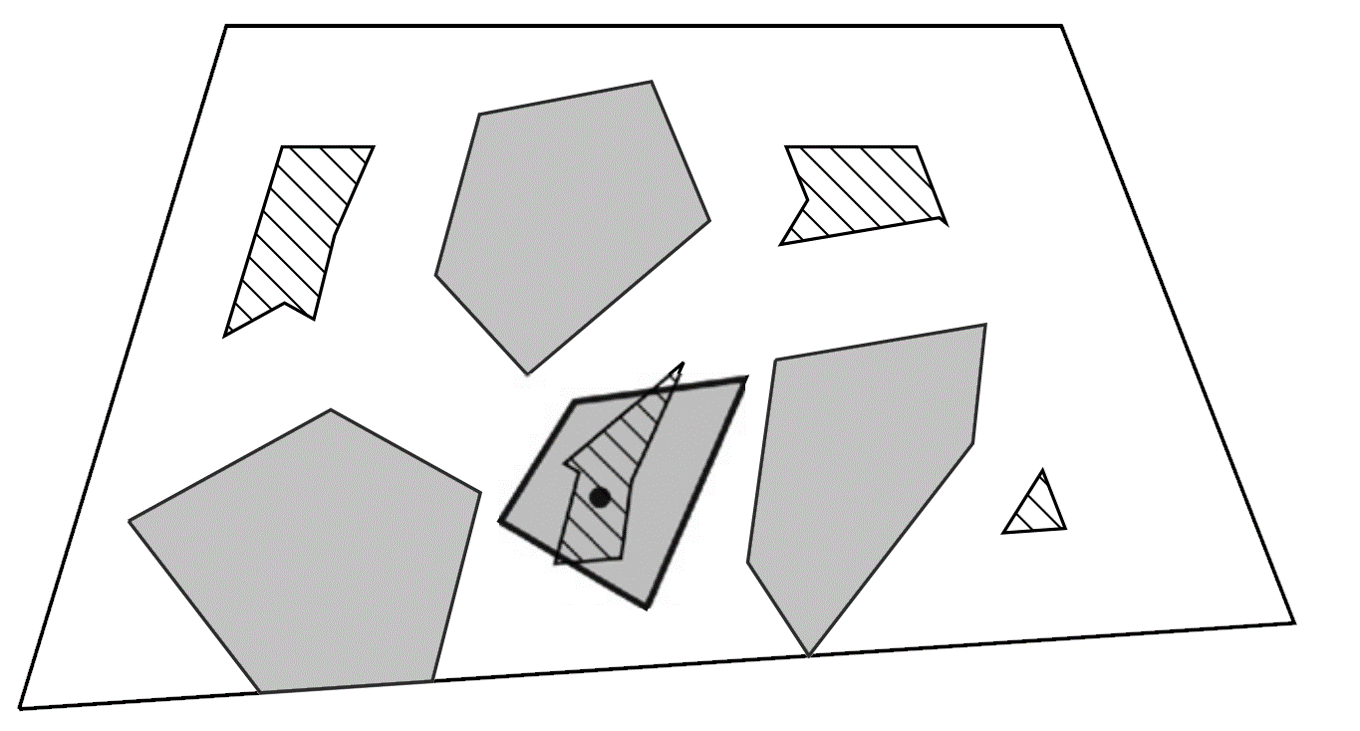}
\caption{\label{fig:NFP_CFR}Colision Free Region example.(\citep{kn:Tsuzuki2013})}
\end{figure}

The approach based on the generation of a Collision Free Region (\citep{kn:Tsuzuki2013}) is created through boolean operations involving the NFP and IFP. This approach is still not widely used but it is very relevant. Its main advantage is the representation through a single polygonal structure that defines the CFR instead of using multiple simpler polygons (as when using convex decomposition) which simplifies overlap verification computations. The main downside arises when concavities and degenerated cases appear which make overlap verifications much more complex and computational expensive. An example of the CFR can be seen in Fig.~\ref{fig:NFP_CFR}.

The NFP is an efficient method to compute efficiently the relative position between pairs of pieces but the processes available for its generation are still limited. The correct identification of holes and perfect fit or perfect sliding placement positions need is still a computational expensive operation (due to being treated as degenerate cases, among other reasons) and there are still many problems derived from numerical precision which cause errors that need to be addressed. 

The next section presents a new approach that aims to address some of these challenges.

\section{Proposed NFP Algorithm} 
\label{sec:method}

In this section is presented a simple algorithm that reduces the challenges generating NFPs that contain holes, perfect sliding placement positions and perfect fit placement positions. As a side not, this algorithm can also be used to merge normal polygons, however, the perfect fit and perfect sliding placement positions will be discarded because normal polygons do not have zero area regions.

\subsection{Convex Decomposition and NFP Generation}

The convex decomposition of the pieces is achieved through any algorithm using partition or covering decomposition approaches. As explained in the example of Polygonal Convex Decomposition, one of the main concerns is the total number of sub-polygons and vertices generated, which impact significantly and proportionally the computational cost. In this algorithm the number of sub-polygons and vertices is not considered a significant concern since the complexity of the final NFP generated will not have any dependence on it. The number of vertices and convex sub-polygons will only impact the computation times to produce the NFP in a pre-processing phase, which may be discarded in most practical applications (since its done only once).

\begin{figure}[!htbp]
\centering
\includegraphics[scale=0.325]{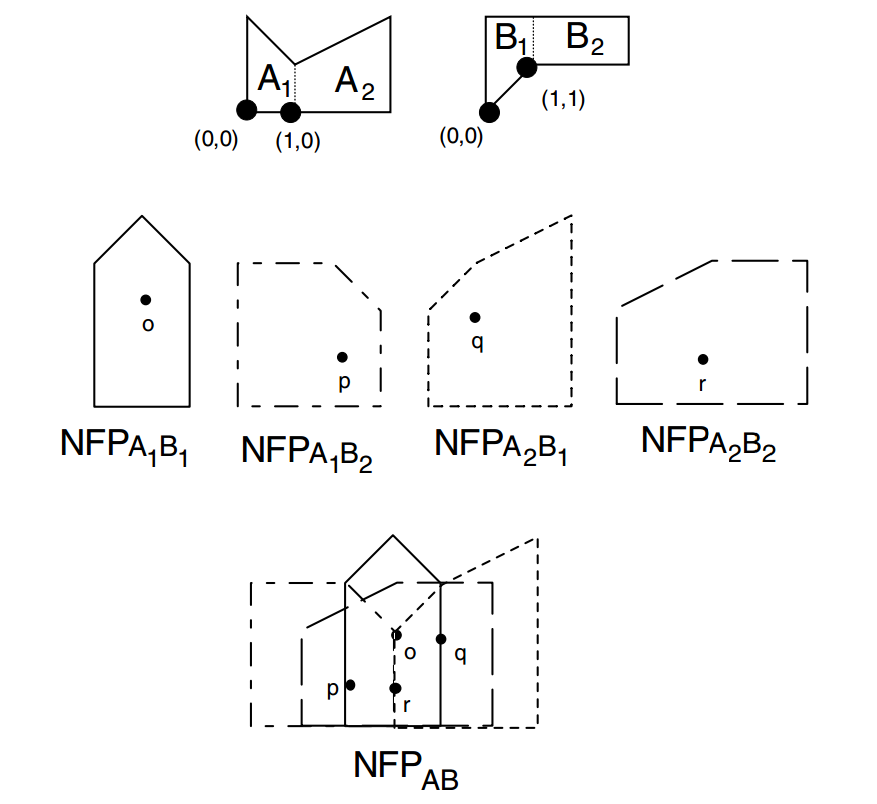}
\caption{\label{fig:NFP_Dcmp}NFP generated by convex decomposition.(\citep{kn:Bennell2008a})}
\end{figure}

The convex NFP generation algorithm uses Slope Diagrams based approach similar to the ones present in \citep{kn:Bennell2008a} and in \citep{kn:Bennell2008b}. One example can be seen in Fig.~\ref{fig:NFP_Dcmp}.

\subsection{NFP Algorithm Demonstration}

\begin{figure}[!htbp]
\centering
	\subfloat[PA][Orbital Piece $A$\label{subfig:PieceA}]{
   		\includegraphics[scale=0.20]{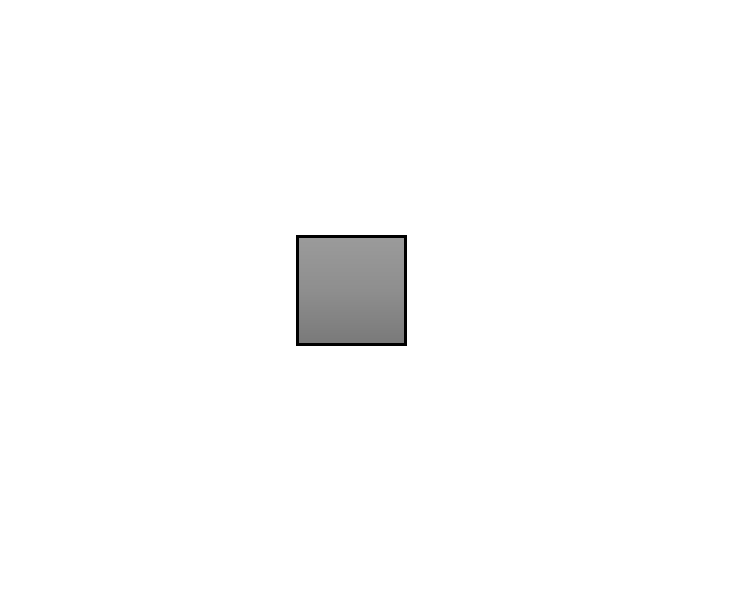} 
 	}
 	\subfloat[PB][Stationary Piece $B$\label{subfig:PieceB}]{
   		\includegraphics[scale=0.20]{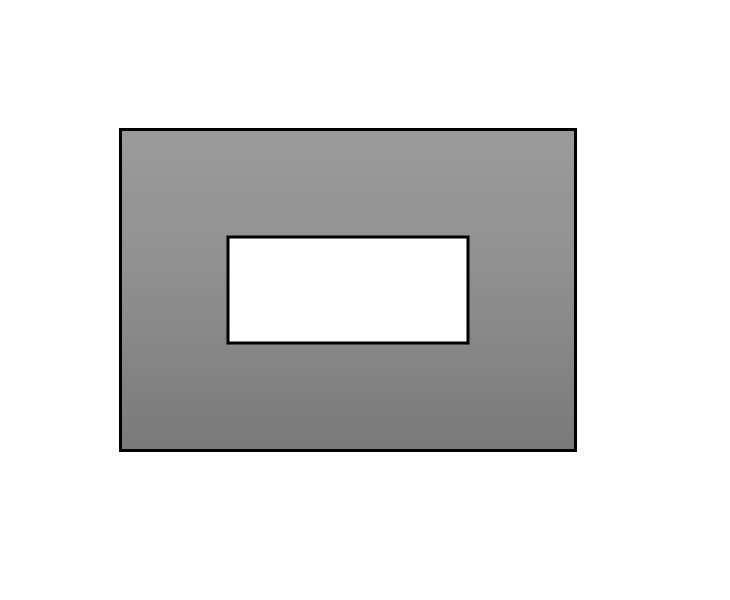}   
 	}    
\caption{Pieces $A$ and $B$ used to generate $NFP_{AB}$\label{fig:figPAB}}
\end{figure}

The process to generate the full NFP between two different pieces $A$ and $B$, which can be seen in Fig.~\ref{fig:NFP_Dcmp}, is demonstrated below. The pieces must be convex or be defined as a convex set of components. This is achieved using an algorithm for convex decomposition that will generate convex components of the original pieces. In the example of Fig.~\ref{fig:figPAB}, piece $B$ is already defined as a set of convex components.

\begin{figure}[!htbp]
\centering
	\subfloat[N2][Pieces $A$ and $B$\label{subfig:NFP_Example2all}]{
   		\includegraphics[scale=0.20]{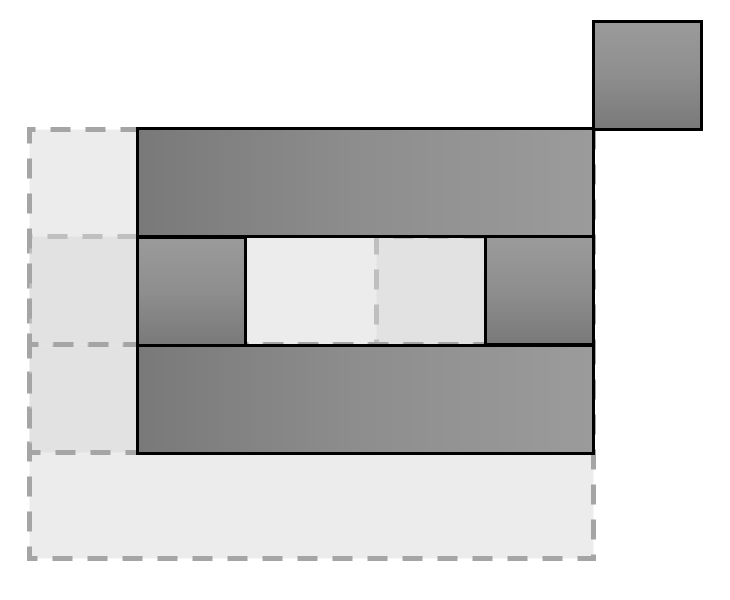} 
 	}
 	\subfloat[N3][$NFP_{AB}$\label{subfig:NFP_Example3}]{
   		\includegraphics[scale=0.20]{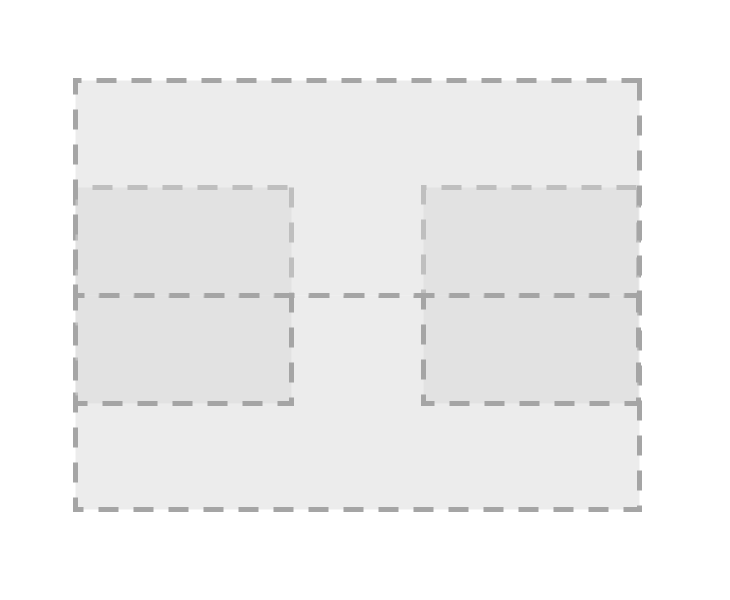}   
 	}
    \subfloat[N4][Intersections\label{subfig:NFP_Example4}]{
   		\includegraphics[scale=0.20]{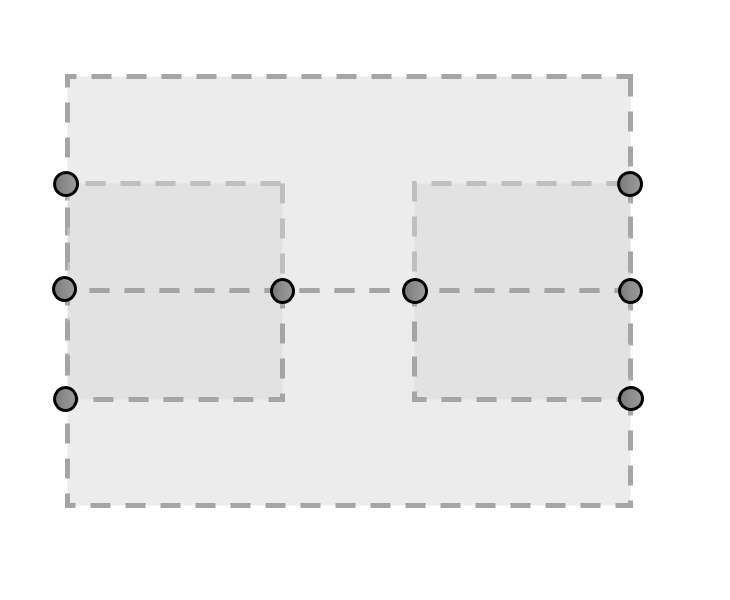}   
 	}
\caption{Generating convex $NFP_{AB}$ intersections\label{fig:figNFP234}}
\end{figure}

The first step is to produce the $NFP_{AB}$ components derived from all pairs of convex components between pieces $A$ and $B$, as seen in Fig.~\ref{fig:figNFP234}(a), and (b). Then, intersections between pairs of NFP components can be computed, and merged into each respective NFP that generated them, as seen in Fig.~\ref{fig:figNFP234}(c).

\begin{figure}[!htbp]
\centering
	\subfloat[N2][Convex $NFP_{AB}$ vertices (with intersections)\label{subfig:NFP_Example5}]{
   		\includegraphics[scale=0.20]{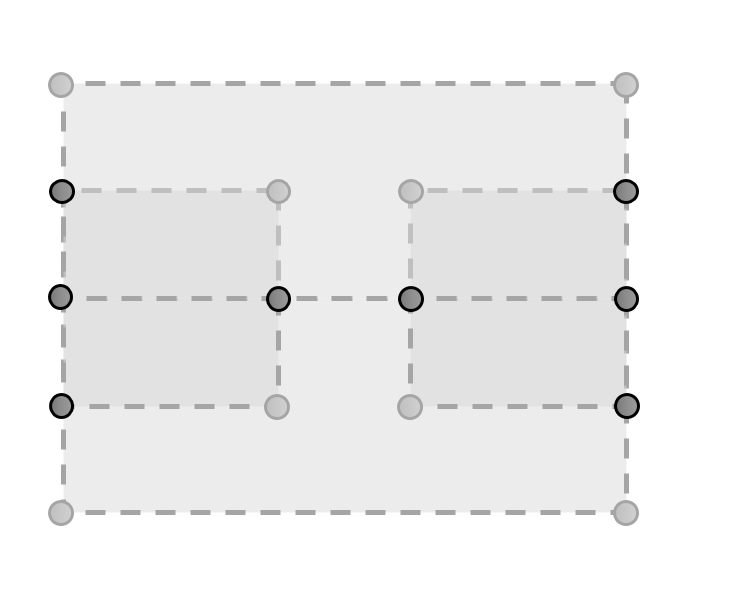} 
 	}
    \quad
    \quad
 	\subfloat[N3][Convex $NFP_{AB}$ vertices (with intersections and edge midpoints)\label{subfig:NFP_Example6}]{
   		\includegraphics[scale=0.20]{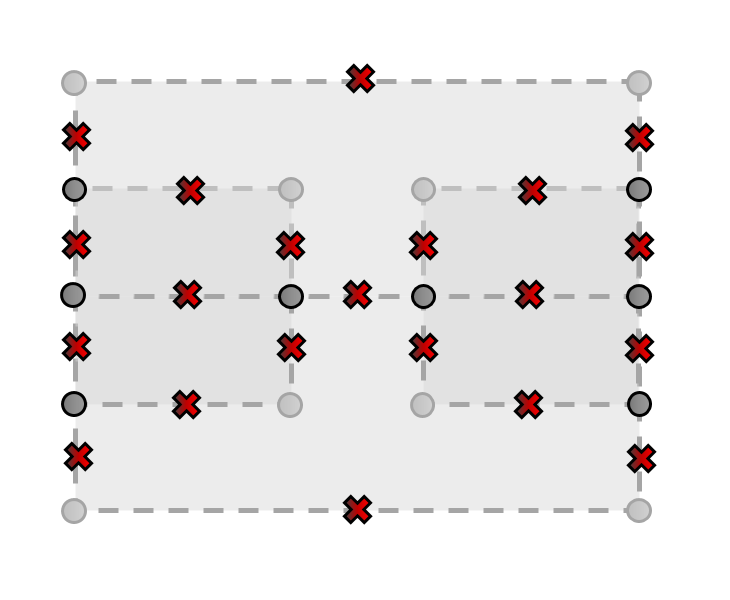}   
 	}   
\caption{$NFP_{AB}$ intersections\label{fig:figNFP56}}
\end{figure}

Any vertices that are derived from intersections are integrated into the respective NFP outlines, producing the structure seen in Fig.~\ref{fig:figNFP56}(a), with all original vertices of the NFP components and all the intersection vertices. The following step is a division of all of the NFP component segments in half, by their midpoints, generating a new vertex that is integrated into the respective NFP components, as seen in Fig.~\ref{fig:figNFP56}(b).

The convex NFP components are then converted into a directed graph, with all vertices (original, intersections and midpoints) being placed in their positions with the respective connections (edges) between each pair, in the same direction that was used to define the interior of each convex NFP component (i.e. counter-clockwise direction). 

With this conversion into a graph, the similar vertices are aggregated together and their connections to other vertices are defined. In situations of multiple identical links between vertices (due to conversion into graph), only one link (or edge) is introduced into the graph.

\begin{figure}[!htbp]
\centering
	\subfloat[N2][Removal of all vertices contained inside the convex NFPs\label{subfig:NFP_Example7}]{
   		\includegraphics[scale=0.20]{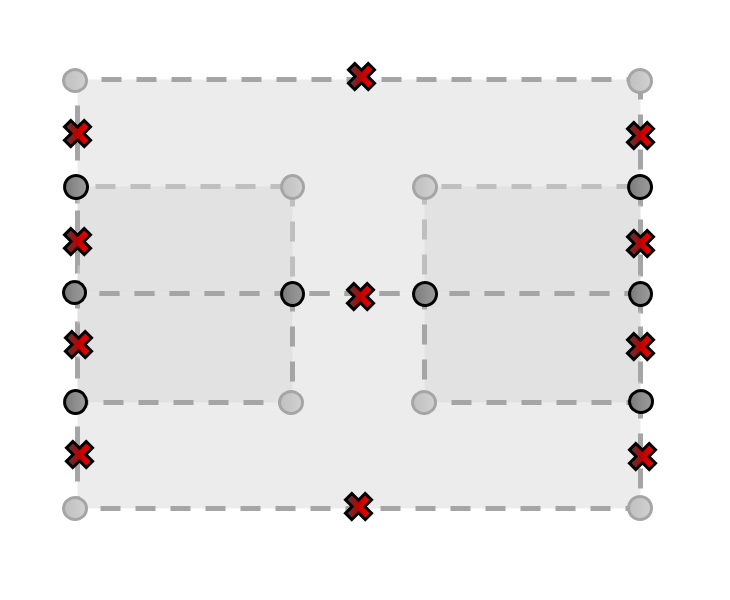} 
 	}
    \quad
    \quad
 	\subfloat[N3][Extraction of NFP outline, and identification of holes\label{subfig:NFP_Example8}]{
   		\includegraphics[scale=0.20]{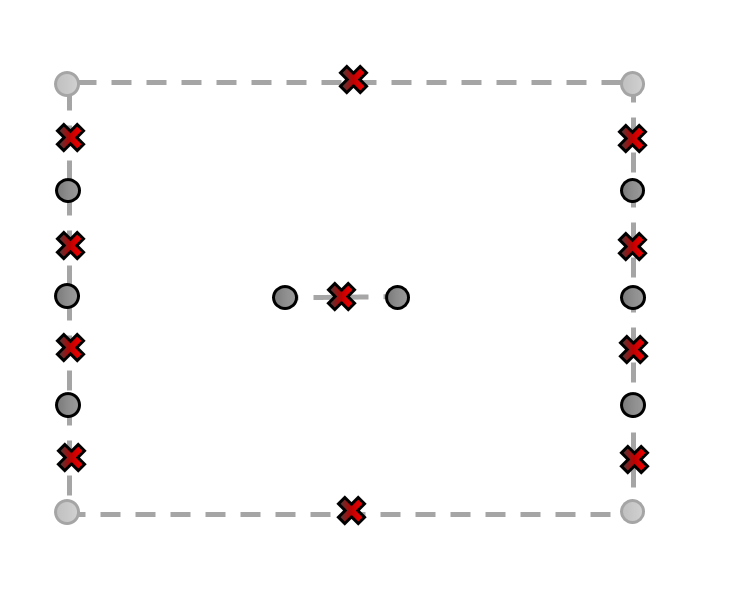}   
 	}   
\caption{$NFP_{AB}$ outlines and holes\label{fig:figNFP78}}
\end{figure}

Every vertex is then verified for containment inside each convex NFP component, and if containment is verified (and not intersecting the outline) the vertex is removed, and all edges that connect to it are deleted. This process removes all internal vertices, leaving only the vertices on the outline of the NFP components that form the full NFP, including holes, perfect sliding and perfect fit placement locations, as shown in Fig.~\ref{fig:figNFP78}(a) and (b). This step produces the full NFP, but it does not discriminate between the different components. In order to correctly identify outlines, holes and perfect sliding and fit locations, another procedure must be done.

\begin{figure}[!htbp]
\centering
	\subfloat[First combination][Outline of $NFP_{AB}$\label{subfig:NFP_Example9a}]{
   		\includegraphics[scale=0.20]{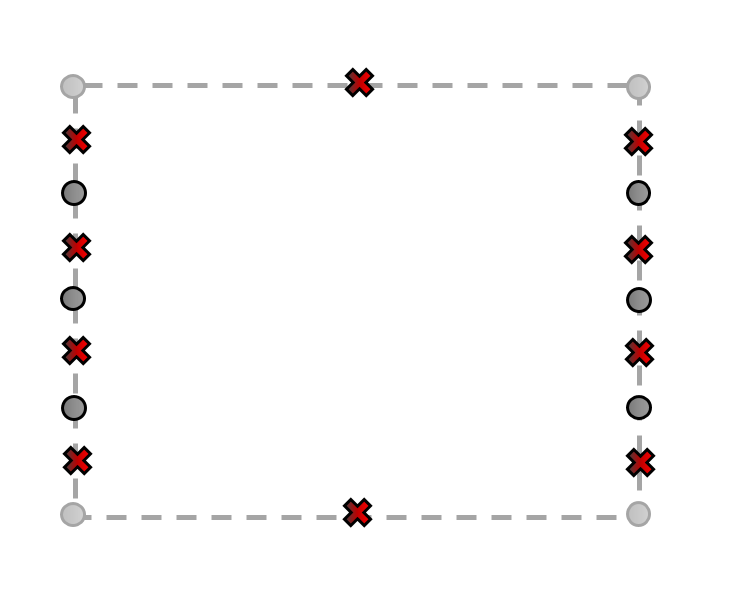} 
 	}
    \quad
    \quad
 	\subfloat[Second combination][Hole of $NFP_{AB}$ perfect slide\label{subfig:NFP_Example9b}]{
   		\includegraphics[scale=0.20]{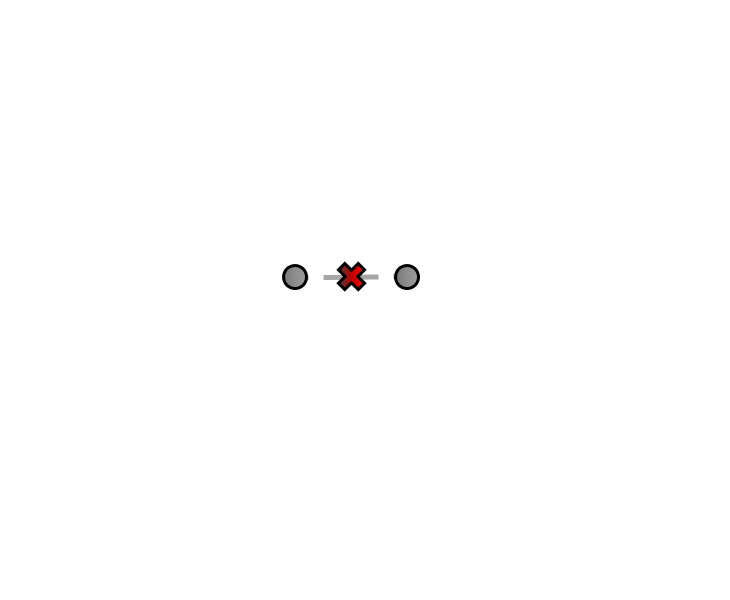}   
 	}   
\caption{Generating $NFP_{AB}$ outline and holes\label{fig:figNFP9ab}}
\end{figure}

To extract the different components, an algorithm that finds the most exterior outlines is used. This is explained below in section~\ref{sec:sss_oea} and exemplified in section~\ref{sec:ss_oee}. It identifies each outline, hole and perfect sliding or perfect fit locations, as seen in Fig.~\ref{fig:figNFP9ab}, and returns them individually.

The resulting graph represents the full NFP between original polygons. The last step is to convert the graph into the polygonal representation of the NFP with its holes, perfect fit and perfect sliding locations (identified by isolated vertices or sub-circuits without area). In order to identify holes or sub-circuits, one needs to use a specific method to detect exterior paths. 

\subsubsection{Outline Extraction Algorithm}
\label{sec:sss_oea}

The algorithm used to extract the outlines of the pieces starts by selecting, among the vertices of the graph $G_{AB}$, one or multiple vertices with the lowest coordinates in the X-axis, and among those, the single vertex with minimum y-axis coordinates. This vertex, $V$, is the starting point from which the outline will be extracted. 

\begin{figure}[!htbp]
\centering
\includegraphics[scale=0.235]{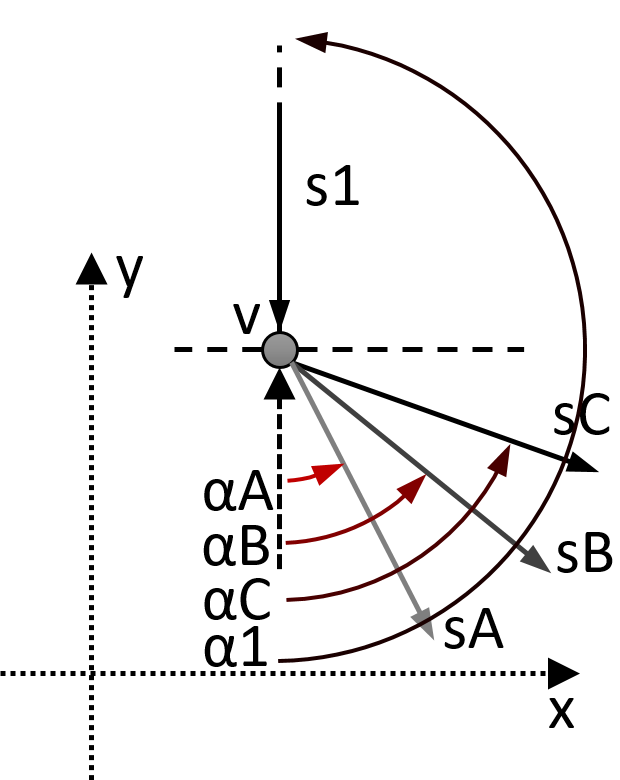}
\caption{\label{fig:PathExtract5}Selection of edge linked to $V$ with smallest angle relative to vertical segment}
\end{figure}

From $V$, all edges connected to it are compared considering their angle relative to a vertical segment connected to $V$. The edge(s) with the smallest angle relative to this vertical segment are selected and their direction is verified (Inward or Outward). As an example, in Fig.~\ref{fig:PathExtract5} there is only one edge connected to $V$ with the smallest angle ($S_A$) given by $\alpha_A$. Considering the same approach, the edge with the maximum angle relative to the vertical segment linked to $V$ is $S_1$ with the angle $\alpha_1$. 

\begin{figure}[!htbp]
\centering
	\subfloat[P001][Outward edge $S_2$\label{subfig:PathExtract1}]{
   		\includegraphics[scale=0.235]{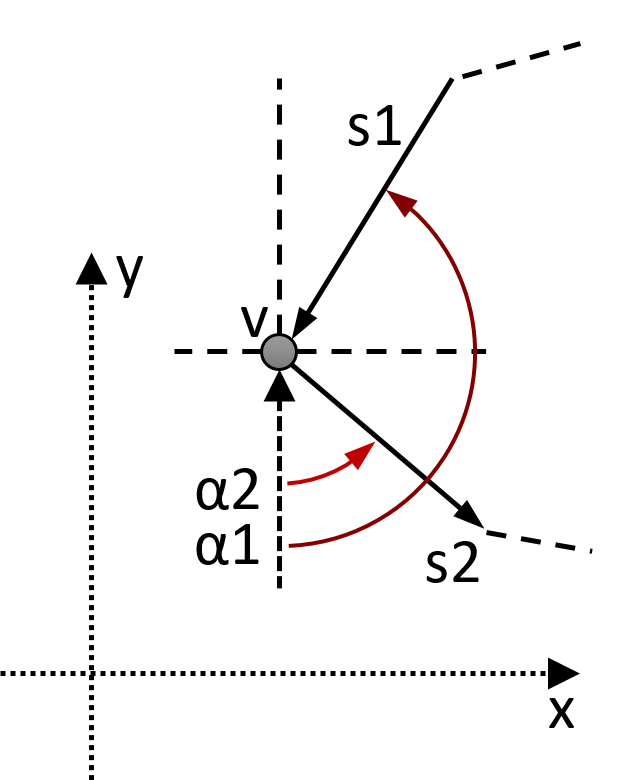} 
 	}
 	\subfloat[P002][Inward edge $S_A$\label{subfig:PathExtract2}]{
   		\includegraphics[scale=0.235]{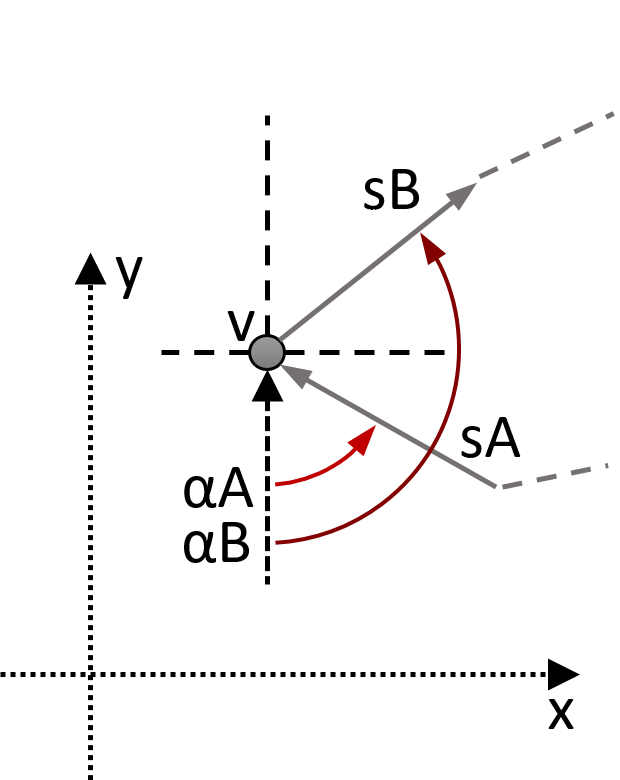}   
 	}    
    \subfloat[P003][Outline and hole at $V$\label{subfig:PathExtract3}]{
   		\includegraphics[scale=0.235]{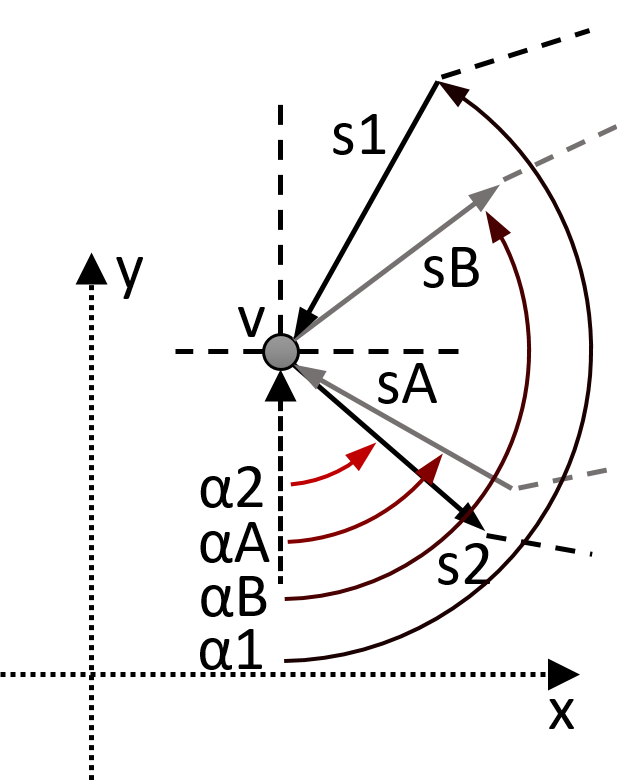} 
 	} 	
    \subfloat[P003][Perfect sliding hole\label{subfig:PathExtract4}]{
   		\includegraphics[scale=0.235]{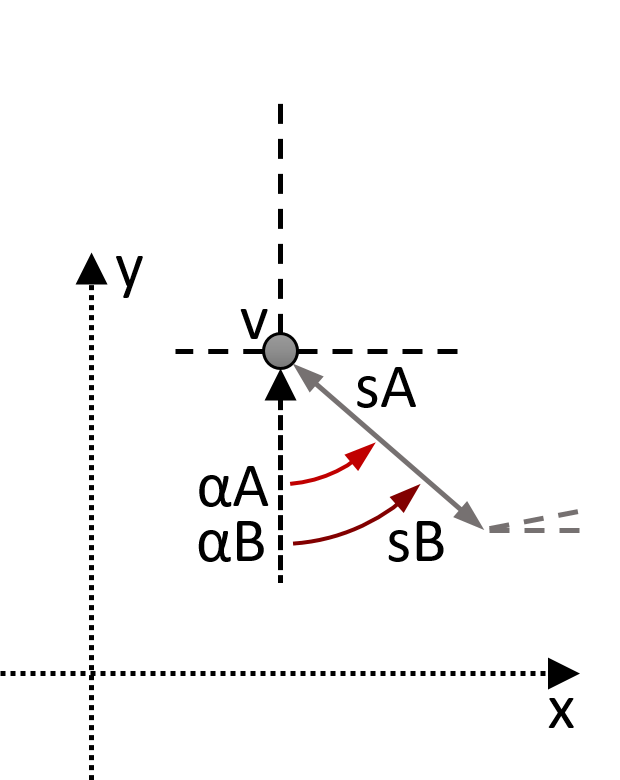} 
 	} 	
\caption{Edge selection for outline and hole extraction\label{fig:figPE1234}}
\end{figure}

The type of structure (outline, hole) can be deduced from the direction of the edge with minimum relative angle to the vertical segment. Four different cases can be presented, as shown in Fig.~\ref{fig:figPE1234}.

In the first case, in \ref{subfig:PathExtract1}, the edge ($S_2$) has an outward direction, which defines an outline, and not a hole. The outline starts at $V$ and progresses through the edges until ending at $V$. In \ref{subfig:PathExtract2}, the edge ($S_A$) has an inward direction, which defines a hole that starts at $V$ and progresses through the edges in reverse direction also ending at $V$. 

Considering case \ref{subfig:PathExtract3}, $V$ contains both an outline and a hole. The edge that is selected ($S_2$) defines an outline, which is also extracted and finishes at $V$ from edge ($S_1$). With the outline extracted, its edges are removed and the process starts over again. Vertex $V$ is selected again, but this time, edge $S_A$ is selected, defining a hole, which is then also extracted. 

The distinct behavior occurs in \ref{subfig:PathExtract4}, where there are 2 edges connected to $V$ that end and start at the same vertex, due to having contrary directions, inward and outward. In this case, the structure is defined as a perfect sliding hole. If a vertex without connections is found (in the initial phase when searching for the vertex with minimum x-axis and y-axis coordinate) then this structure is defined as a perfect fit hole.

The resulting structure is the complete NFP with minimal geometric components (vertices and edges).

\subsubsection{Outline Extraction Example}
\label{sec:ss_oee}
\begin{figure}[!htbp]
\centering
	\subfloat[P001][NFP example\label{subfig:PathExt1}]{
   		\includegraphics[scale=0.225]{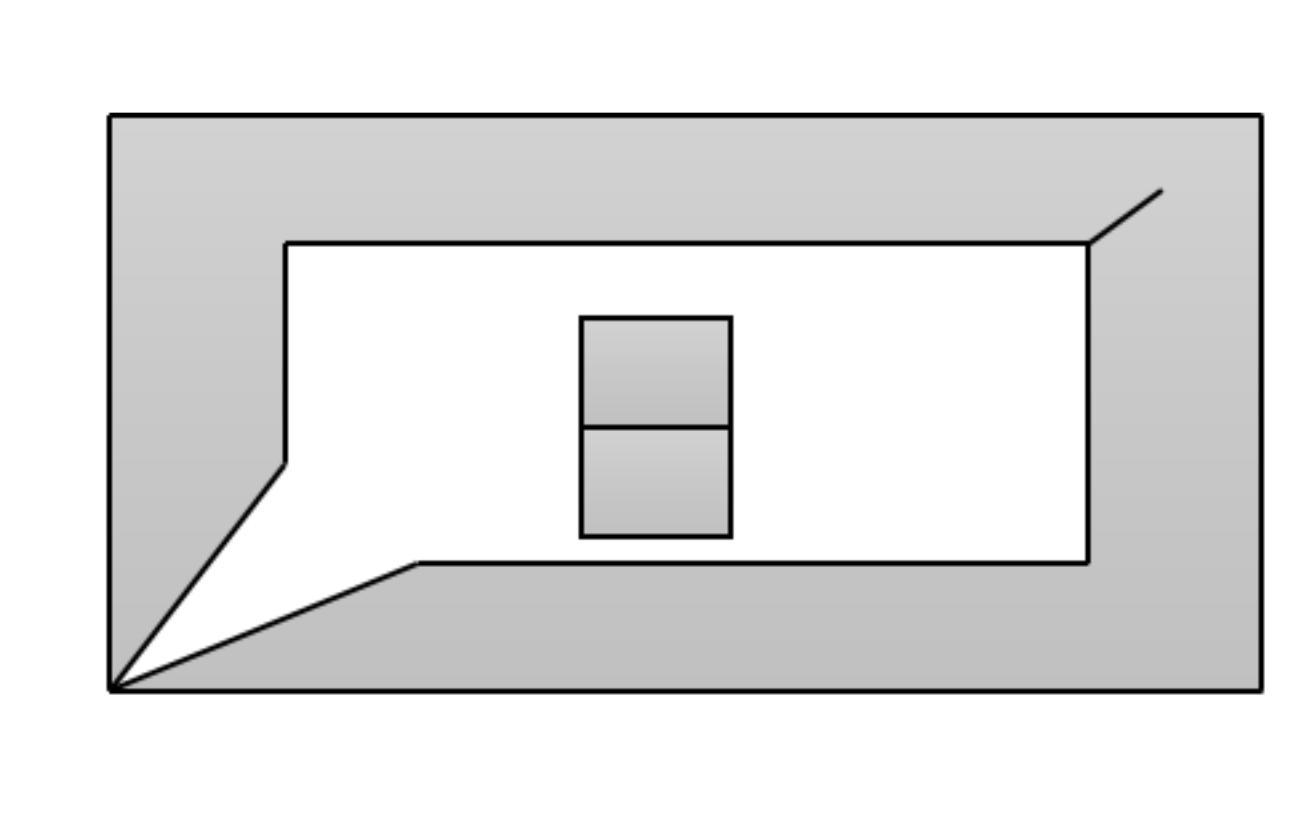} 
 	}
 	\subfloat[P002][NFP directed edges\label{subfig:PathExt2}]{
   		\includegraphics[scale=0.225]{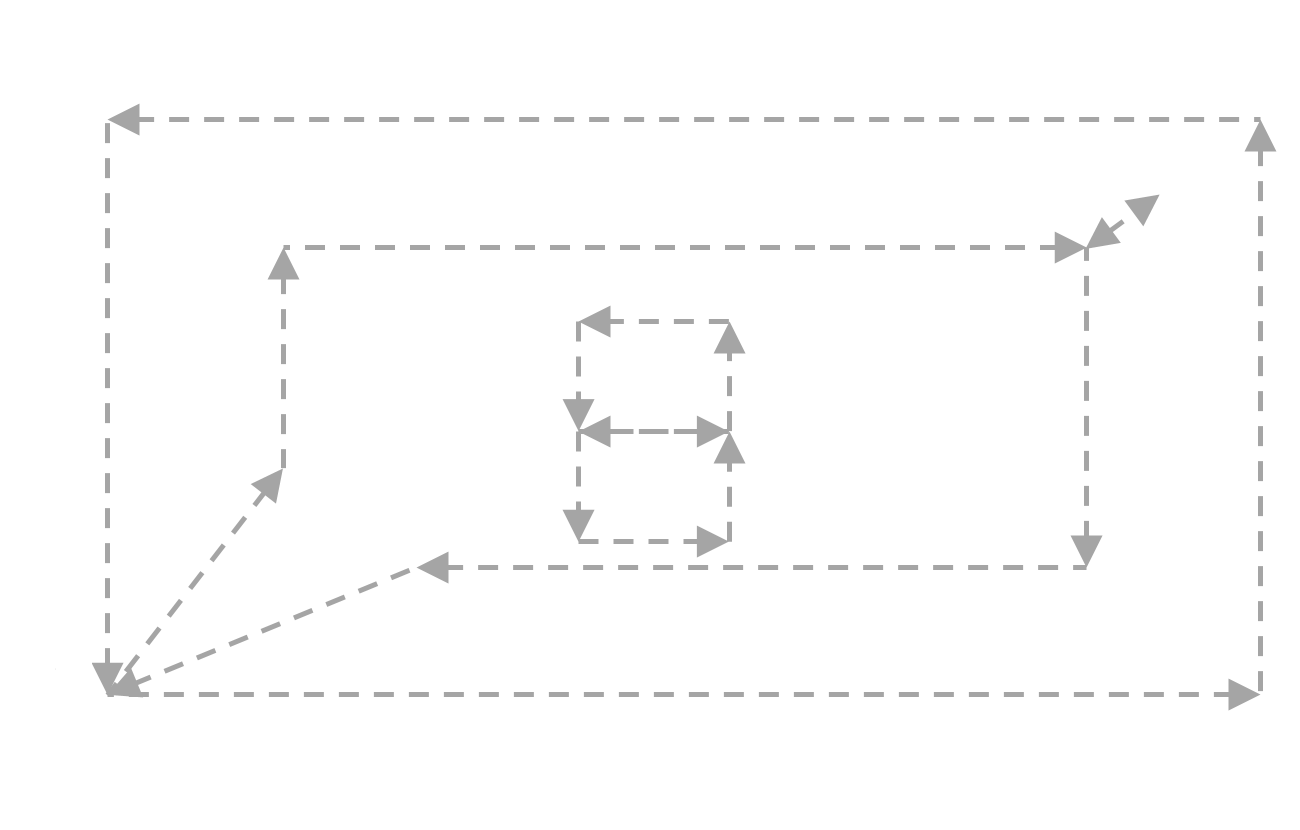}   
 	}    
    
    \subfloat[P003][First circuit detected (outline)\label{subfig:PathExt3}]{
   		\includegraphics[scale=0.225]{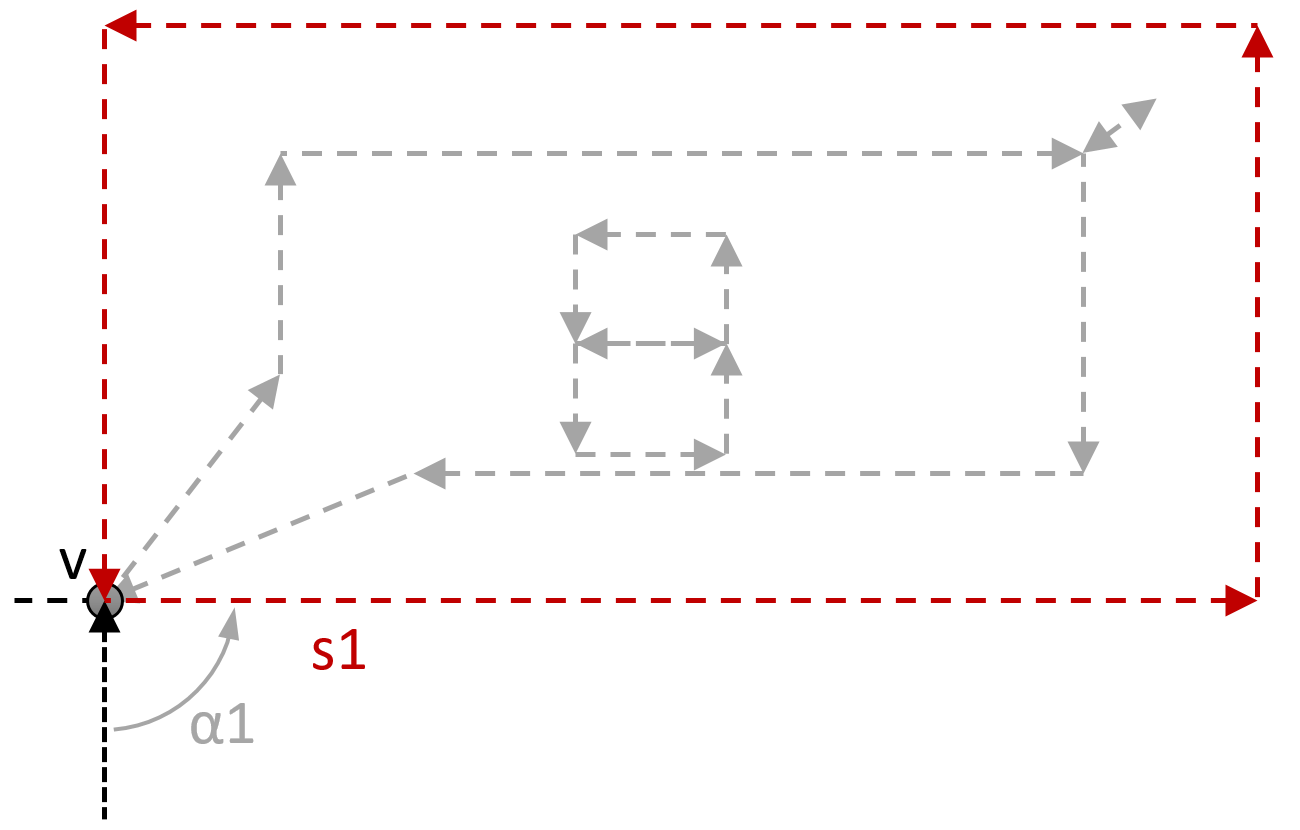} 
 	} 	
    \subfloat[P004][Second circuit detected (hole)\label{subfig:PathExt4}]{
   		\includegraphics[scale=0.225]{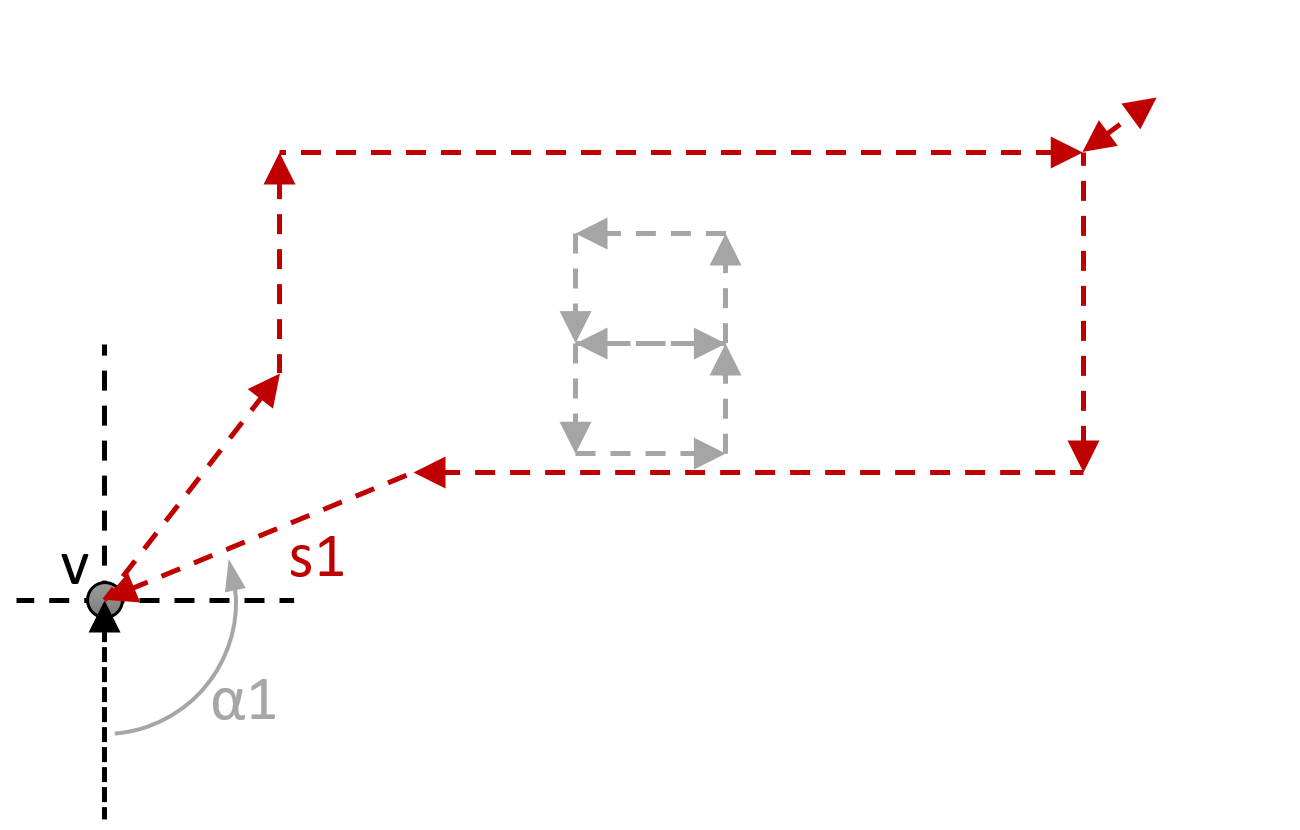} 
 	} 	
    
    \subfloat[P005][Third circuit detected (outline)\label{subfig:PathExt5}]{
   		\includegraphics[scale=0.225]{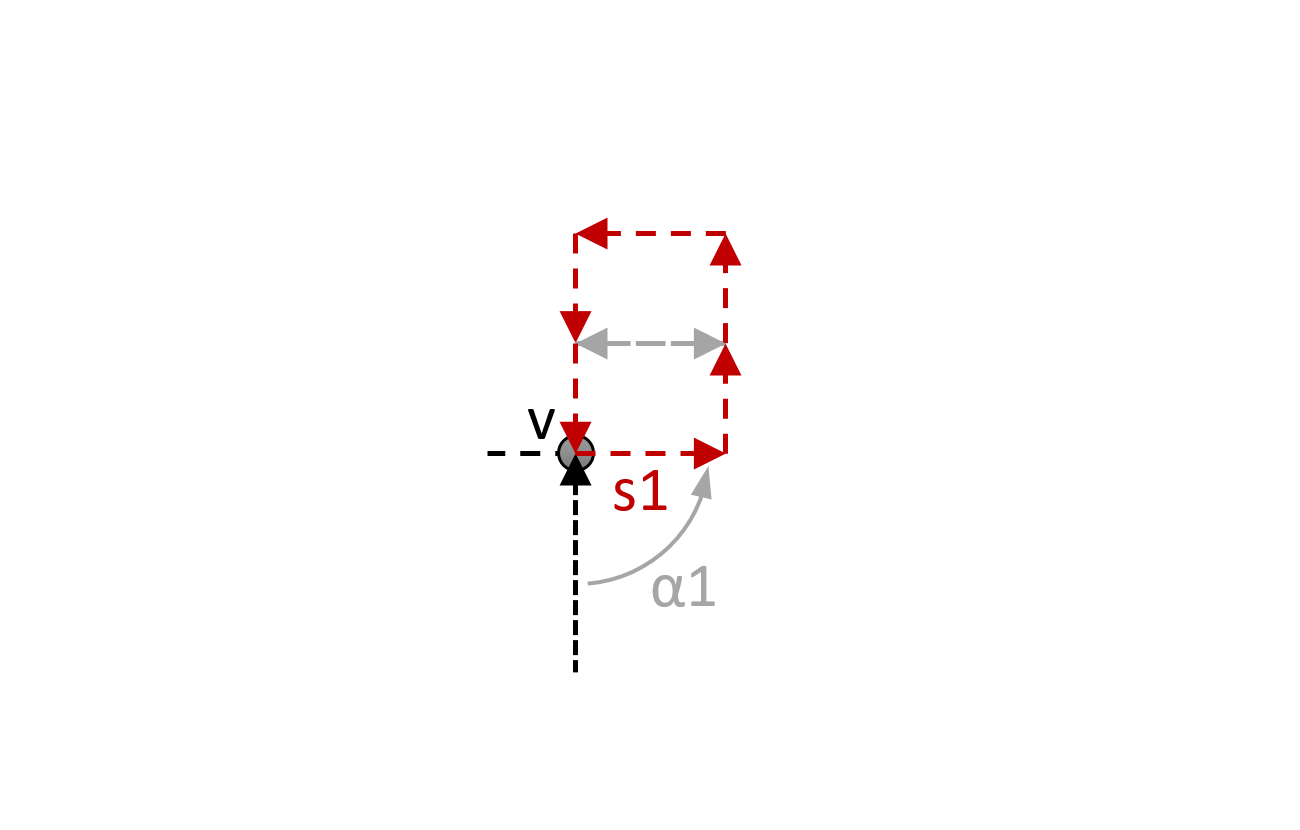} 
 	} 	
    \subfloat[P006][Fourth circuit detected (hole)\label{subfig:PathExt6}]{
   		\includegraphics[scale=0.225]{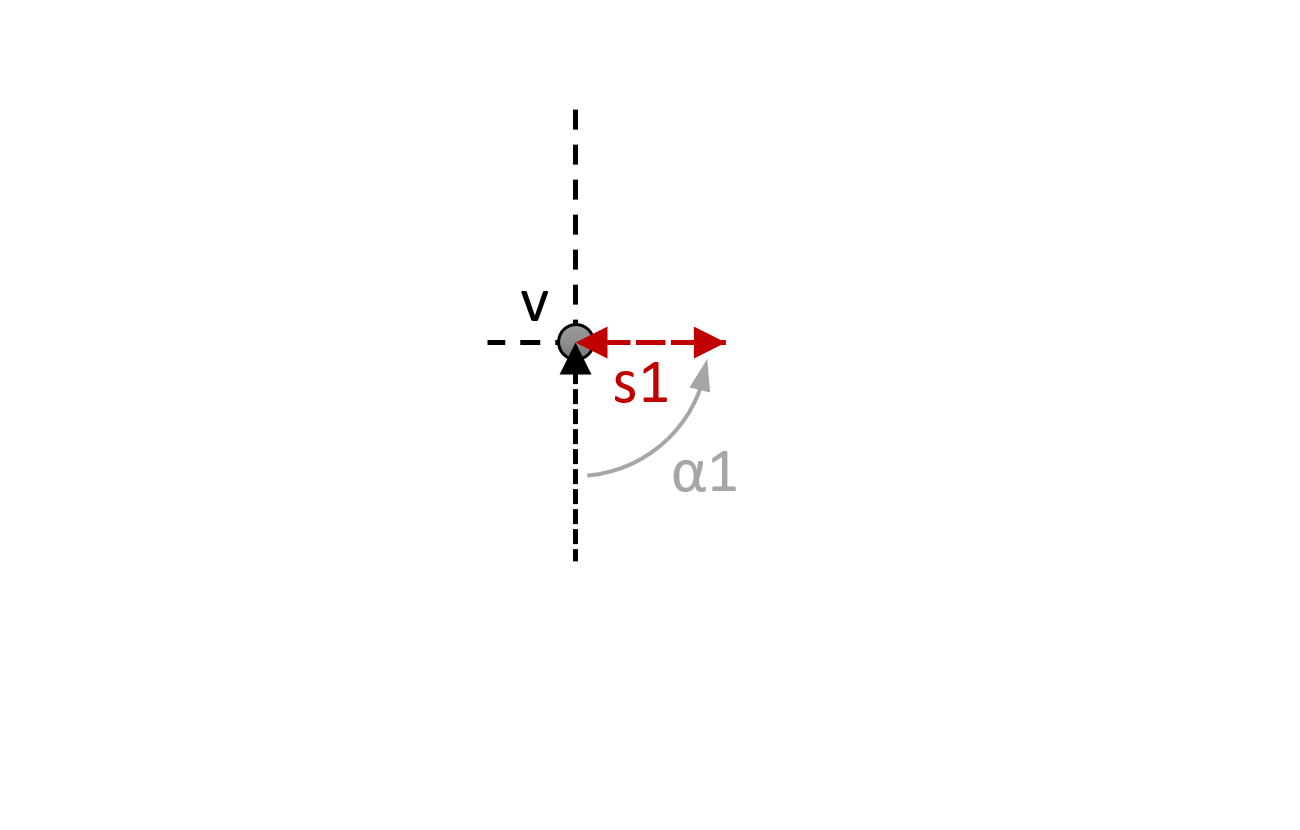} 
 	} 	
\caption{Outline and Hole Extraction Demonstration\label{fig:figPE123456}}
\end{figure}

The outline and hole extraction process is presented in Fig~\ref{fig:figPE123456}. The NFP shown in Fig.~\ref{subfig:PathExt1} is defined by its directed edges (Fig.~\ref{subfig:PathExt2}). It starts by selecting the vertices with the lowest coordinates in the X-axis, and among those, the single vertex with minimum y-axis coordinates (shown as $V$ in Fig.~\ref{subfig:PathExt3}). From $V$, the edge with smallest relative angle to a vertical segment is selected and its direction is verified (Inward or Outward). This distinction allows defining if the most exterior circuit defines a normal outline or a hole. In Fig.~\ref{subfig:PathExt3} the outward edge $S_1$ with minimum angle $\alpha_1$ is selected and the outline is extracted (in red).
The next iteration, in Fig.~\ref{subfig:PathExt4}, repeats this process by finding $V$ and $S_1$ through the minimum angle $\alpha_1$, and the next circuit is extracted (in this case since $S_1$ is Inward, the circuit is a hole).
Fig.~\ref{subfig:PathExt5}, repeats the process identifying an outline while in Fig.~\ref{subfig:PathExt6} a zero area hole is extracted.

\subsection{Algorithm Description}
\label{sec:ss_nma}

The proposed algorithm generates the full NFP using convex decomposition of the original pieces, generating the NFP from the combinations of the convex sub-polygons, and merging all of them together while correctly identifying holes, perfect slide locations and perfect fit locations in the layout.

The general steps of the NFP generation process is shown in the pseudo-code in Algorithm~\ref{gennfpalgo}: 
\begin{algorithm}
	\caption{NFP Generation Algorithm}
    \label{gennfpalgo}
	\begin{algorithmic}[1]
     \Procedure{GENNFP}{$A,B$} 														\Comment{The $NFP_{AB}$ of pieces $A$ and $B$}
    	\State {Convex decomposition for pieces $A$ and $B$} 						\Comment{Covering or partition algorithm}	\label{lst:line:al1}
        \State {Compute $NFP_{AB}$ between convex sub-polygons}						\Comment{Using slope diagrams}				\label{lst:line:al2}
        \State {Compute intersection points between $NFP_{AB}$ components}			\Comment{All edge pairs intersections}		\label{lst:line:al3}
        \State {Merge intersection points into respective $NFP_{AB}$ component}		\Comment{All edge pairs intersections}		\label{lst:line:al4}
        \For {Each $NFP_{AB}$ component edge}									
        	\State {Compute and merge midpoint vertex}   							\Comment{Divide edges by midpoint}			\label{lst:line:al5}
        \EndFor
        \State {Create graph $G_{AB}$ from $NFP_{AB}$ components}					\Comment{Define graph by linking common vertices} \label{lst:line:al6}
  		\For {Each $NFP_{AB}$ component $C$}										
        	\For {Each $NFP_{AB}$ vertex $V$}
            	\If{$V$ contained inside $C$}
                    \State {Eliminate $V$ from $G_{AB}$} 							\Comment{Eliminate any contained vertex}	\label{lst:line:al7}
        			\State {Eliminate edges linked to $V$ from $G_{AB}$}            \Comment{Eliminate linked edges}			\label{lst:line:al8}
                \EndIf
            \EndFor          	             
  		\EndFor  	 
        \State {Rebuild $NFP_{AB}$ from $G_{AB}$* using OExAlgo}     \Comment{Outlines, holes, perfect slide and fit positions} \label{lst:line:al9}      
        \State \textbf{Return} $NFP_{AB}$
        \EndProcedure
        \State {*Using Outline Extraction Algorithm described in Algorithm~\ref{outextalgo}}	
	\end{algorithmic}
\end{algorithm}

The NFP Generation Algorithm (GENNFP) starts by applying convex decomposition of both pieces $A$ and $B$ (in line~\ref{lst:line:al1}) and proceeding to generate all NFP combinations between the pairs of convex components of the respective pieces $A$ and $B$ (line~\ref{lst:line:al2}). These two steps are the pre-requisite for the correct execution of the algorithm. The following step (line~\ref{lst:line:al3}) consists of computing the intersections between all generated NFPs, and integrating these vertices into the respective NFPs (line~\ref{lst:line:al4}). The segment of each NFP is then divided by its midpoint, by integrating a dividing vertex in the middle of the segments of the NFP (line~\ref{lst:line:al5}). In order to assist the next steps, all NFPs are converted into a graph structure, with directed edges, and shared common vertices between different NFPs (line~\ref{lst:line:al6}). The vertices are then tested for containment inside each convex NFP, and the vertices found inside (and not on the outline) are then removed, together with all edges that connect to it ((line~\ref{lst:line:al7} and line~\ref{lst:line:al8}). The final step of the algorithm is the identification and extraction of the outlines and holes (including perfect sliding and perfect fit locations), which is then returned as the complete NFP (line~\ref{lst:line:al9}). This is done using another algorithm that is presented in the pseudo-code in Algorithm~\ref{outextalgo}.

\begin{algorithm}
    \caption{Outline Extraction Algorithm}
    \label{outextalgo}
    \begin{algorithmic}[1] 
    \Procedure{OExAlgo}{$G_{AB}$} 													\Comment{The $NFP_{AB}$ from $G_{AB}$}
        	
            \State {Get set $V_{minx}$ of vertices with the minimum X-axis coordinate from $G_{AB}$} 						\label{lst:line:bl1}
            \State {Select vertex $V$ with the minimum Y-axis coordinate from set $V_{minx}$} 							\label{lst:line:bl2}
            
            \State {Select edges $E$ linked to vertex $V$ with minimum angle relative to a upward vertical segment} 	\label{lst:line:bl3}
                       
            \If{$E$ has one inward edge}																				\label{lst:line:bl4}
            	\State {Extract hole (may be a perfect sliding)} 					\Comment{Outward path starting at $E$}               
            \ElsIf{$E$ has only outward edge}																			\label{lst:line:bl5}
                \State {Extract outline} 											\Comment{Outward path starting at $E$}
            \ElsIf{$E$ has no connecting edges}																			\label{lst:line:bl6}
                \State {Extract hole ($V$ is a perfect fit)}						\Comment{Hole $V$}
            \EndIf
                            
            \State \textbf{Return} extracted objects (outlines, holes)										\label{lst:line:bl7}
            
        \EndProcedure
    \end{algorithmic}
\end{algorithm}

The outline extraction algorithm starts by selecting the vertex that has the minimum X-axis coordinate (line~\ref{lst:line:bl1}), and also the minimum y-axis coordinate (line~\ref{lst:line:bl2}), in this order of preference. This ensures that the vertex that is selected will not have any other vertex with a lower x-axis coordinate and, among similar vertices that have the same x-axis coordinate will select one ($V$) with the minimum y-axis coordinate. The step in (line~\ref{lst:line:bl3}) selects the edge or edges (connected to $V$) that have the smallest relative angle to a vertical segment linked to $V$. If $V$ has two edges (with minimum angle) then it selects the edge with inward direction. If it only has one edge with inward direction then it is a normal hole. This is verified at line~\ref{lst:line:bl4}. If the edge selected has an outward direction (line~\ref{lst:line:bl5}), then an outline is extracted. If a vertex is found initially without any connections, then its defined as a perfect fit hole (line~\ref{lst:line:bl6}). During the extraction process, all vertices joining collinear edges are discarded. The objects are then returned in line~\ref{lst:line:bl7}, and the algorithm ends.

\section{Validation using Literature Problem Cases} 

The NFPs shown in the following Fig.~\ref{fig:NFP_brk} were obtained using the current algorithm using the respective pieces and parameters as described in \citep{kn:Burke2007}. The piece coordinates were obtained using high detail rasterization of their figures. The NFPs were correctly generated and all degenerate cases correctly identified. 

The first problem case shown in Fig.~\ref{subfig:burke1} deals with a pair of interlocking concavities that produce a NFP with a single hole. This case requires the detection of feasible placement positions derived from interactions of the concavities to generate holes in the NFP. The detection of holes is easier than other degenerate cases due to holes having empty regions of space with area. The larger the hole, the easier it is to detect it using different strategies due to larger range of motion possible for the placement of a piece. Detecting a perfect fit is much harder since it requires detecting a placement position without overlap between pieces. Holes may not be detected when using orbital approaches without adequate methods to detect feasible placement positions that define the holes.

In Fig.~\ref{subfig:burke2} is presented a problem case with multiple interlocking concavities. This case uses two similar pieces, where one is rotated $180^{\circ}$, and is a typical problem case for orbital approaches in the literature (as in \citep{kn:Mahadevan1984}). Due to its geometrical characteristics, when tackled by orbital approaches it requires the identification of multiple starting points created by the interactions of the concavities of both pieces. The holes have multiple different sizes, so some of the approaches may be able to detect the larger ones, but not the smallest. The current algorithm was able to successfully identify all five internal holes and the external outline.

Fig.~\ref{subfig:burke3} presents a problem case where perfect fit sliding positions are available. This problem is challenging due to difficulties determining feasible sliding translation movements. This requires the identification of exact placement positions, although with some range of movement across a certain direction. This is a challenge due to numerical precision and algorithm limitations.

The problem case of Fig.~\ref{subfig:burke4} presents an exact fit between jigsaw pieces (where the pieces are locked together without one of them being contained inside the other and unable to move). The significant challenge of this problem case is the requirement of detecting the exact placement position without any gap or range of motion. It requires dealing adequately with any numerical precision problems that arise. 

The last problem case, in Fig.~\ref{subfig:burke5} also deals with the challenges of detecting feasible starting positions involving holes. One of the holes of the stationary piece has multiple starting positions leading to a NFP where several distinct holes are present. The pieces used in this case are of significantly higher complexity than the ones used in previous problem cases. There is a mix of holes and interlocking concavities. If the approach used to address this case uses convex decomposition, the number of polygons generated may significantly increase its difficulty, due to the large number of NFPs produced (either by the high computational cost of the aggregate NFP set, or by the challenge of having an algorithm able to merge the convex NFPs to reconstruct the full NFP).

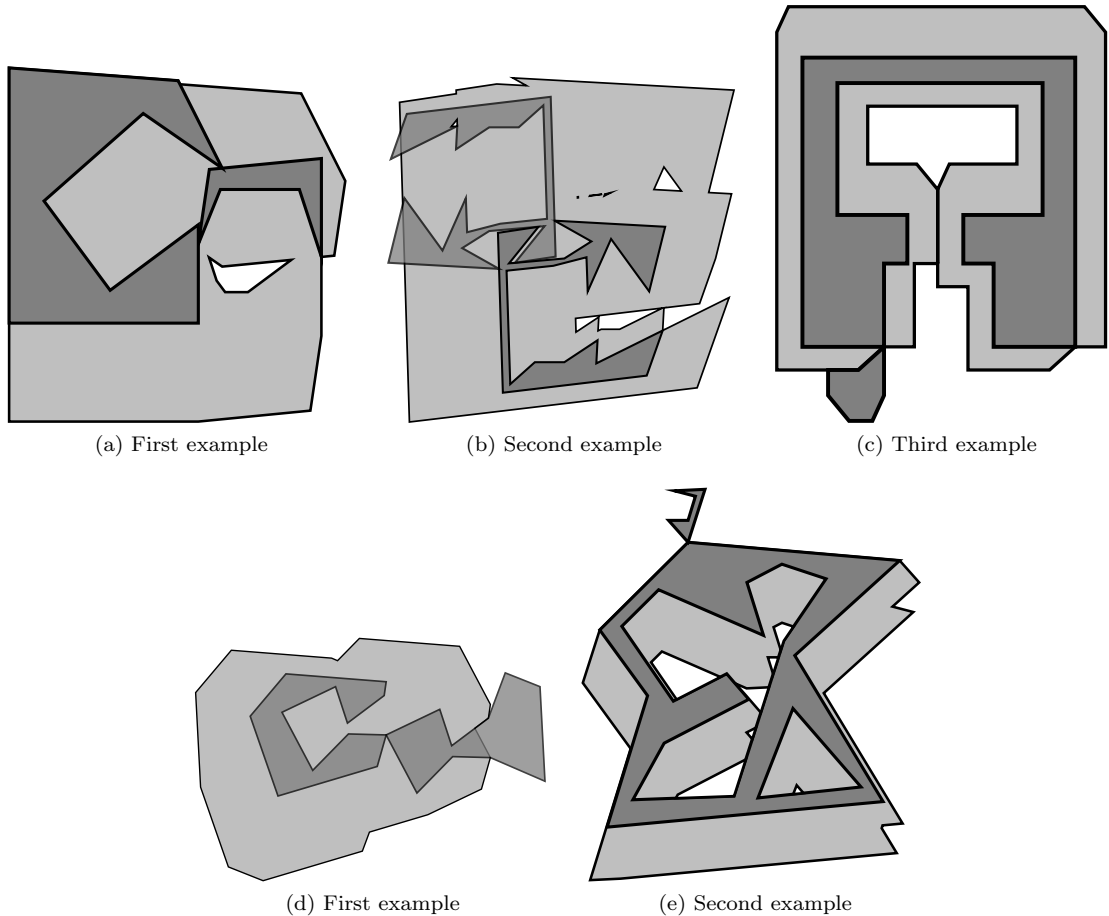
\begin{figure}[!htbp]
\centering
  \subfloat[First combination][First example\label{subfig:burke1}]{    
    \resizebox{0.30\linewidth}{!}{
      	\begin{tikzpicture}[scale=0.05]

\draw			[ very thick,line width=1pt,fill=lightgray, shift={(00,00)}, opacity=1.0]	
(	0	,	0	) --
(	51.5	,	0	) --
(	82	,	3	) --
(	85	,	23.5	) --
(	85	,	45.1557	) --
(	88.5	,	45.5	) --
(	91.5	,	66	) --
(	79.5	,	90	) --
(	46.4944	,	92.5113	) --
(	46	,	93.5	) --
(	0	,	97	) --
cycle;			
	
\draw				[very thick,line width=1pt,fill=white, shift={(00,00)}, opacity=1.0]
(	54.4196	,	44.998	) --
(	58	,	42.5	) --
(	76.7987	,	44.3491	) --
(	65	,	35.5	) --
(	58.7959	,	35.5	) --
(	56.4758	,	38.6579	) --
cycle;			
	
\draw				[very thick,fill=gray, shift={(00,00)}, opacity=1.0]
(	51.5	,	48.6557	) --
(	57.5	,	63.6557	) --
(	79	,	63.6557	) --
(	85	,	45.1557	) --
(	85	,	72.1557	) --
(	54.5	,	69.1557	) --
cycle;			

\draw				[very thick,fill=gray, shift={(00,00)}, opacity=1.0]
(	0	,	97	) --
(	0	,	27	) --
(	51.5	,	27	) --
(	51.5	,	54	) --
(	27.5	,	36	) --
(	9.5	,	60.5	) --
(	36.5	,	84.5	) --
(	58	,	69.5	) --
(	46	,	93.5	) --
cycle;				

\end{tikzpicture}
    }
  }
  \subfloat[Second combination][Second example\label{subfig:burke2}]{
    \resizebox{0.30\linewidth}{!}{
    	\begin{tikzpicture}[scale=0.05]

\draw				[ very thick,line width=1pt,fill=lightgray, shift={(00,00)}, opacity=1.0]	
(	4.55	,	131	) --
(	8.55	,	0	) --
(	125.55	,	14	) --
(	138.55	,	51	) --
(	111.55	,	37.5	) --
(	112.0376	,	46.7635	) --
(	126.55	,	48.5	) --
(	133.05	,	67	) --
(	139.55	,	93.5	) --
(	130.2522	,	94.0165	) --
(	140.55	,	136	) --
(	50.55	,	141	) --
(	56.1478	,	137.8279	) --
(	44.05	,	138.5	) --
(	27.55	,	136	) --
(	27.596	,	134.4918	) --
cycle;		

\draw				[very thick,fill=gray, shift={(00,00)}, opacity=1.0]
(	44.55	,	77.5	) --
(	46.55	,	12	) --
(	105.05	,	19	) --
(	111.55	,	37.5	) --
(	84.55	,	24	) --
(	85.05	,	33.5	) --
(	71.55	,	24.5	) --
(	59.55	,	24.5	) --
(	49.55	,	15.5	) --
(	48.05	,	62	) --
(	67.05	,	64	) --
(	80.55	,	67.5	) --
(	81.05	,	53.5	) --
(	90.55	,	75	) --
(	106.05	,	53.5	) --
(	112.55	,	80	) --
(	67.55	,	82.5	) --
(	82.55	,	74	) --
(	71.55	,	67	) --
(	49.05	,	65	) --
(	61.05	,	80	) --
cycle;	

\draw				[very thick,fill=gray, shift={(00,00)}, opacity=0.75]
(	68	,	67.75	) --
(	66	,	133.25	) --
(	7.5	,	126.25	) --
(	1	,	107.75	) --
(	28	,	121.25	) --
(	27.5	,	111.75	) --
(	41	,	120.75	) --
(	53	,	120.75	) --
(	63	,	129.75	) --
(	64.5	,	83.25	) --
(	45.5	,	81.25	) --
(	32	,	77.75	) --
(	31.5	,	91.75	) --
(	22	,	70.25	) --
(	6.5	,	91.75	) --
(	0	,	65.25	) --
(	45	,	62.75	) --
(	30	,	71.25	) --
(	41	,	78.25	) --
(	63.5	,	80.25	) --
(	51.5	,	65.25	) --
cycle;				

\draw			[ very thick,line width=1pt,fill=white, shift={(00,00)}, opacity=1.0]		
(	25.55	,	121	) --
(	27.918	,	123.9596	) --
(	28.001	,	121.2179	) --
cycle;		

\draw				[ very thick,line width=1pt,fill=white, shift={(00,00)}, opacity=1.0]	
(	76.1156	,	42.4651	) --
(	111.432	,	46.691	) --
(	94.05	,	38	) --
(	86.55	,	38	) --
(	85.2527	,	37.3514	) --
(	85.55	,	43	) --
(	76.2974	,	36.8316	) --
cycle;		

\draw				[ very thick,line width=1pt,fill=white, shift={(00,00)}, opacity=1.0]	
(	76.8799	,	92.2716	) --
(	77.2176	,	92.3071	) --
(	77.2335	,	91.8611	) --
(	76.8948	,	91.8098	) --
cycle;	

\draw				[ very thick,line width=1pt,fill=white, shift={(00,00)}, opacity=1.0]	
(	81.769	,	92.5483	) --
(	81.8793	,	92.7978	) --
(	86.9121	,	93.3276	) --
cycle;	

\draw				[ very thick,line width=1pt,fill=white, shift={(00,00)}, opacity=1.0]	
(	87.7303	,	92.5902	) --
(	88.0308	,	93.4453	) --
(	88.55	,	93.5	) --
(	90.6269	,	94.0385	) --
cycle;		

\draw				[ very thick,line width=1pt,fill=white, shift={(00,00)}, opacity=1.0]	
(	107.9649	,	95.2547	) --
(	112.05	,	104.5	) --
(	119.1637	,	94.6326	) --
cycle;				
	
\end{tikzpicture}
    }
  } 
  \subfloat[Third combination][Third example\label{subfig:burke3}]{
  	\resizebox{0.30\linewidth}{!}{
   		\begin{tikzpicture}[scale=0.05]

\draw				[ very thick,line width=1pt,fill=lightgray, shift={(00,00)}, opacity=1.0]
(	0	,	11	) --
(	17.5	,	11	) --
(	23	,	16	) --
(	29.5	,	16	) --
(	29.5	,	34	) --
(	34.5	,	34	) --
(	34.5	,	29	) --
(	41	,	29	) --
(	41	,	11	) --
(	58.5	,	11	) --
(	64	,	16	) --
(	70.5	,	16	) --
(	70.5	,	84	) --
(	66	,	89.5	) --
(	2.5	,	89.5	) --
(	0	,	84	) --
cycle;			

\draw				[very thick,line width=1pt,fill=white, shift={(00,00)}, opacity=1.0]
(	19.5	,	55.5	) --
(	19.5	,	68	) --
(	51.5	,	68	) --
(	51.5	,	55.5	) --
(	37	,	55.5	) --
(	34.5	,	50	) --
(	34.5	,	34	) --
(	34.5	,	50	) --
(	30	,	55.5	) --
cycle;		

\draw				[very thick,fill=gray, shift={(00,00)}, opacity=1.0]
(	5.5	,	78.5	) --
(	5.5	,	16	) --
(	23	,	16	) --
(	23	,	34	) --
(	28	,	34	) --
(	28	,	44.5	) --
(	13	,	44.5	) --
(	13	,	73	) --
(	57	,	73	) --
(	57	,	44.5	) --
(	40	,	44.5	) --
(	40	,	34	) --
(	46.5	,	34	) --
(	46.5	,	16	) --
(	64	,	16	) --
(	64	,	78.5	) --
cycle;				

\draw				[very thick,fill=gray, shift={(00,00)}, opacity=1.0]
(	11	,	11	) --
(	11	,	5.5	) --
(	15.5	,	0	) --
(	20.5	,	0	) --
(	23	,	5.5	) --
(	23	,	16	) --
(	17.5	,	11	) --
cycle;				
		
\end{tikzpicture}
    }
  }
  
  \subfloat[First combination][First example\label{subfig:burke4}]{    
    \resizebox{0.30\linewidth}{!}{
      	\begin{tikzpicture}[scale=0.05]

\draw				 [ very thick,line width=1pt,fill=lightgray, shift={(00,00)}, opacity=1.0]
(	0	,	95.5	) --
(	2.5	,	47.5	) --
(	16.5	,	7	) --
(	34	,	0	) --
(	84	,	15	) --
(	87.57457	,	24.6724	) --
(	117	,	33.5	) --
(	144	,	46.5	) --
(	148.5	,	62.5	) --
(	140.6967	,	77.35152	) --
(	147.5	,	82.5	) --
(	148.5	,	89.5	) --
(	133	,	119	) --
(	82.5	,	123	) --
(	71.4976	,	111.692	) --
(	68.5	,	113	) --
(	18	,	117	) --
cycle;				
\draw				[ very thick,line width=1pt,fill=lightgray, shift={(00,00)}, opacity=1.0]
(	96	,	74	) --
cycle;				
\draw				[very thick,fill=gray, shift={(00,00)}, opacity=0.75]
(	27.5	,	83.5	) --
(	41.5	,	43	) --
(	91.5	,	58	) --
(	96	,	74	) --
(	77	,	74.5	) --
(	59	,	56	) --
(	43.5	,	85.5	) --
(	70.5	,	98.5	) --
(	76.5	,	80	) --
(	95	,	94	) --
(	96	,	101	) --
(	45.5	,	105	) --
cycle;				
\draw				[very thick,fill=gray, shift={(00,00)}, opacity=0.75]
(	96	,	74	) --
(	111.5	,	44.5	) --
(	129.5	,	63	) --
(	148.5	,	62.5	) --
(	176	,	50.5	) --
(	173.5	,	98.5	) --
(	156	,	105.5	) --
(	147.5	,	82.5	) --
(	129	,	68.5	) --
(	123	,	87	) --
cycle;				

\end{tikzpicture}
    }
  }
  \subfloat[Second combination][Second example\label{subfig:burke5}]{
    \resizebox{0.30\linewidth}{!}{
    	\begin{tikzpicture}[scale=0.05]

\draw				[ very thick,line width=1pt,fill=lightgray, shift={(00,00)}, opacity=1.0]
(	0	,	54	) --
(	13	,	36	) --
(	2	,	0	) --
(	10.3	,	0.4	) --
(	85.3	,	7.4	) --
(	81.2107	,	14.215436	) --
(	81.4457	,	14.990936	) --
(	86.9	,	15.5	) --
(	65.4748	,	51.2086	) --
(	89.8	,	73.4	) --
(	84.1507	,	74.8866	) --
(	91.4	,	81.5	) --
(	86.1	,	87.5	) --
(	28.6	,	92.5	) --
(	4.6	,	68.5	) --
cycle;				

\draw				[very thick,fill=gray, shift={(00,00)}, opacity=1.0]
(	57.1	,	47	) --
(	75.6	,	25.5	) --
(	47.6	,	22.5	) --
cycle;		

\draw				[very thick,fill=gray, shift={(00,00)}, opacity=1.0]
(	4.6	,	68.5	) --
(	17.6	,	50.5	) --
(	6.6	,	14.5	) --
(	81.6	,	21.5	) --
(	57.6	,	61.5	) --
(	86.1	,	87.5	) --
(	28.6	,	92.5	) --
cycle;				

\draw				[very thick,fill=lightgray, shift={(00,00)}, opacity=1.0]
(	10.6	,	69.5	) --
(	20.6	,	79.5	) --
(	49.1	,	67	) --
(	44.6	,	81.5	) --
(	54.1	,	86.5	) --
(	66.1	,	82.5	) --
(	54.6	,	65.5	) --
(	41.1	,	23	) --
(	13.6	,	22	) --
(	22.1	,	37.5	) --
(	45.1	,	49.5	) --
(	39.1	,	56.5	) --
(	24.6	,	49	) --
cycle;			

\draw				[very thick,fill=lightgray, shift={(00,00)}, opacity=1.0]
(	57.1	,	47	) --
(	75.6	,	25.5	) --
(	47.6	,	22.5	) --
cycle;		

\draw				[very thick,line width=1pt,fill=white, shift={(00,00)}, opacity=1.0]
(	18.5377	,	59.6377	) --
(	21.4915	,	62.5915	) --
(	37.393	,	55.6171	) --
(	25.4886	,	49.4596	) --
cycle;				
\draw				[very thick,line width=1pt,fill=white, shift={(00,00)}, opacity=1.0]
(	25.2649	,	22.42418	) --
(	25.8	,	23.4	) --
(	44.2917	,	33.0478	) --
(	41.1	,	23	) --
cycle;				
\draw				[very thick,line width=1pt,fill=white, shift={(00,00)}, opacity=1.0]
(	41.4527	,	53.8365	) --
(	44.5	,	52.5	) --
(	50.5634	,	52.7922	) --
(	48.3657	,	45.8734	) --
(	44.4	,	50.5	) --
cycle;				
\draw				[very thick,line width=1pt,fill=white, shift={(00,00)}, opacity=1.0]
(	44.497	,	40.4202	) --
(	47.0579	,	41.7563	) --
(	46.0558	,	38.6016	) --
cycle;				
\draw				[very thick,line width=1pt,fill=white, shift={(00,00)}, opacity=1.0]
(	50.2862	,	61	) --
(	53.1706	,	61	) --
(	51.7116	,	56.407	) --
cycle;				
\draw				[very thick,line width=1pt,fill=white, shift={(00,00)}, opacity=1.0]
(	51.8362	,	69.2612	) --
(	54.1163	,	70.4612	) --
(	57.2496	,	69.4168	) --
(	54.6	,	65.5	) --
(	53.7924	,	62.9577	) --
cycle;				
\draw				[very thick,line width=1pt,fill=white, shift={(00,00)}, opacity=1.0]
(	57.1647	,	23.52479	) --
(	58.1147	,	25.9748	) --
(	59.9647	,	23.82479	) --
cycle;				

\draw				[very thick,fill=gray, shift={(00,00)}, opacity=1.0]
(	24.9	,	106.6	) --
(	30.6	,	105.1	) --
(	28.6	,	98.5	) --
(	23.3	,	98.5	) --
(	28.6	,	92.5	) --
(	33.2	,	107	) --
cycle;				

\end{tikzpicture}
    }
  }
\caption[NFPs and degenerate cases]{\label{fig:NFP_brk} Examples of combinations of polygons that generate NFPs and their degenerated cases (adapted from~\citep{kn:Burke2007})}
    
    The proposed algorithm was able to produce the full NFP for each problem case, detecting all holes and degenerate cases accurately.
\end{figure}

\section{Alternative Applications}

The proposed NFP algorithm presented in section~\ref{sec:method} was primarily developed to deal with the generation of full NFPs in two dimensions, but it can also be used to execute boolean operations between pairs of pieces, with some modifications (an example of this is shown in section~\ref{sec:aa_bo}). These modifications also enable the algorithm to address challenges in a third dimension (as shown in section~\ref{sec:aa_3d}).

\subsection{Boolean Operations}
\label{sec:aa_bo}
\begin{figure}[!htbp]
\centering
	\subfloat[P001][OR (Union)\label{subfig:BoolOP_OR}]{
   		\includegraphics[scale=0.400]{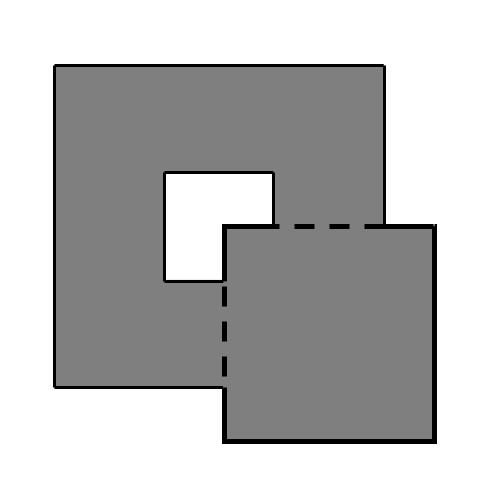} 
 	}
 	\subfloat[P002][AND (Intersection)\label{subfig:BoolOP_AND}]{
   		\includegraphics[scale=0.400]{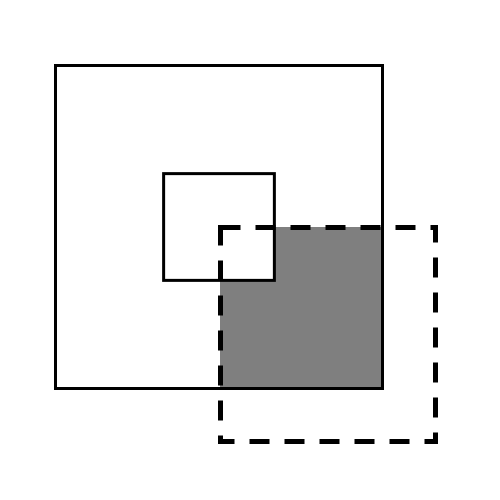}   
 	}    
    
    \subfloat[P003][XOR (Exclusive OR)\label{subfig:BoolOP_XOR}]{
   		\includegraphics[scale=0.400]{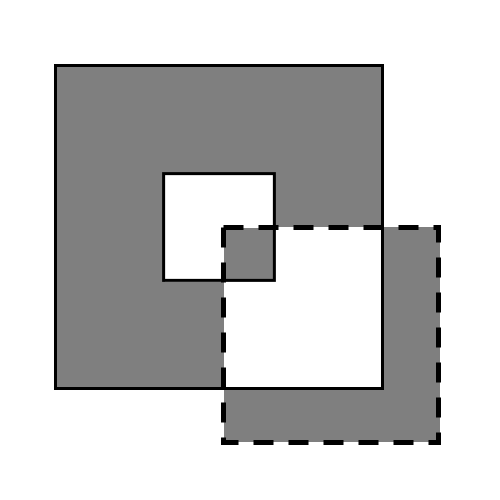} 
 	} 	
    \subfloat[P004][NOT (Negation)\label{subfig:BoolOP_NOT}]{
   		\includegraphics[scale=0.400]{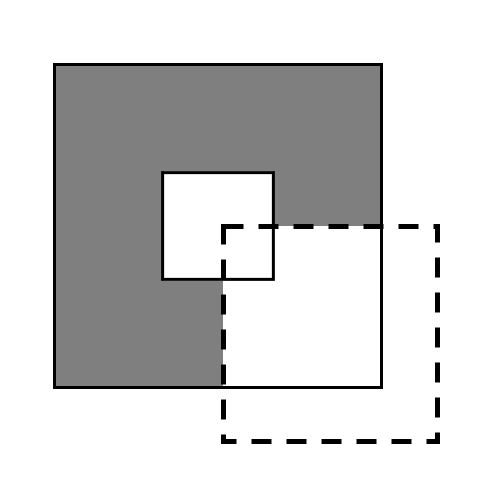} 
 	} 	
\caption{Boolean Operations using different algorithm configuration\label{fig:figBOP}}
\end{figure}

The use of vertices in the middle of each edge allows the computation of boolean operations (such as the ones defined in Fig.~\ref{fig:figBOP}) by selecting vertices to be removed, and edges that will have their direction inverted. Some of the supported boolean operations are intersections (AND), unions (OR), exclusive unions (XOR) and negation (NOT). 

\begin{figure}[!htbp]
\centering
    \subfloat[P005][F1\label{subfig:BoolOP_FIX1}]{
   		\includegraphics[scale=0.400]{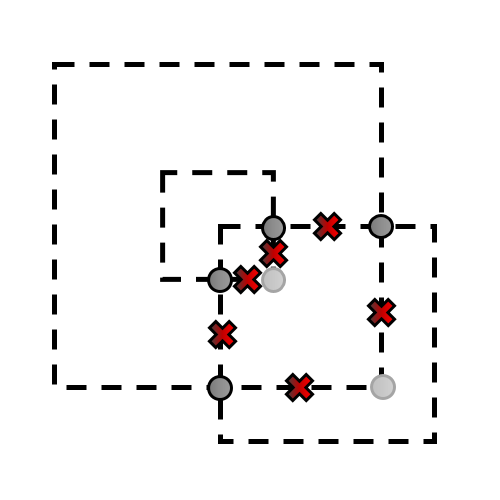} 
 	} 	
    \subfloat[P006][F2\label{subfig:BoolOP_FIX2}]{
   		\includegraphics[scale=0.400]{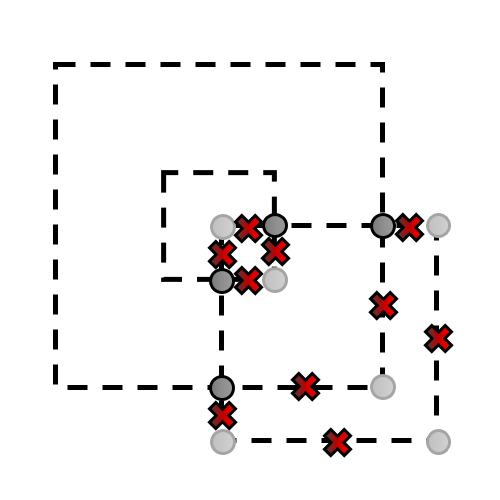} 
 	} 	
\caption{Relevant vertices for boolean operations\label{fig:figBOP2}}
\end{figure}

The boolean operation OR, seen in Fig.~\ref{subfig:BoolOP_OR}, produces the union (or merging) between two pieces. This operation has the same process of the NFP merging algorithm described in this paper, but the zero area regions are ignored (only NFP has those structures). The union of pieces is achieved by removing any vertices that are contained (not on the outline) inside any of the pieces. An example of the vertices that have to be removed can be seen in Fig.~\ref{subfig:BoolOP_FIX1}, where the mid-edge vertices (red) and the outline edges (light gray) are selected. The intersection vertices (dark gray) are not removed.

Regarding the boolean operation AND, seen in Fig.~\ref{subfig:BoolOP_AND}, produces an intersection. This operation follows the same strategy of the boolean operation OR, but with a distinct difference in the selection of the vertices in Fig.~\ref{subfig:BoolOP_FIX1}. In this situation, all vertices are removed except the intersection vertices (dark gray), and the vertices that are inside of at least one of the pieces (mid-edge as red, and outline as light gray).

Considering the operation XOR, seen in Fig.~\ref{subfig:BoolOP_XOR}, the result is achieved not by removing vertices but through inverting the edge directions between vertices. Considering the vertices presented in Fig.~\ref{subfig:BoolOP_FIX1}, all edges connected to any vertex inside either one of the pieces are inverted. Any region that counts as an overlap between the two pieces is now considered a hole.

The boolean operation NOT, in Fig.~\ref{subfig:BoolOP_XOR}, defines a subtraction of one piece by another. In this case the relevant vertices to be removed are are all that are inside the piece that is substracting, the vertices that are maintained (besides the intersecting vertices) are the ones that are inside the piece being subtracted. The vertices that are to be removed are shown in Fig.~\ref{subfig:BoolOP_FIX1}.

\subsection{3D geometry}
\label{sec:aa_3d}

\begin{figure}[!htbp]
\centering
    \subfloat[P005][Before union\label{subfig:3D_BoolOP_OR_Orgnl}]{
   		\includegraphics[scale=0.250]{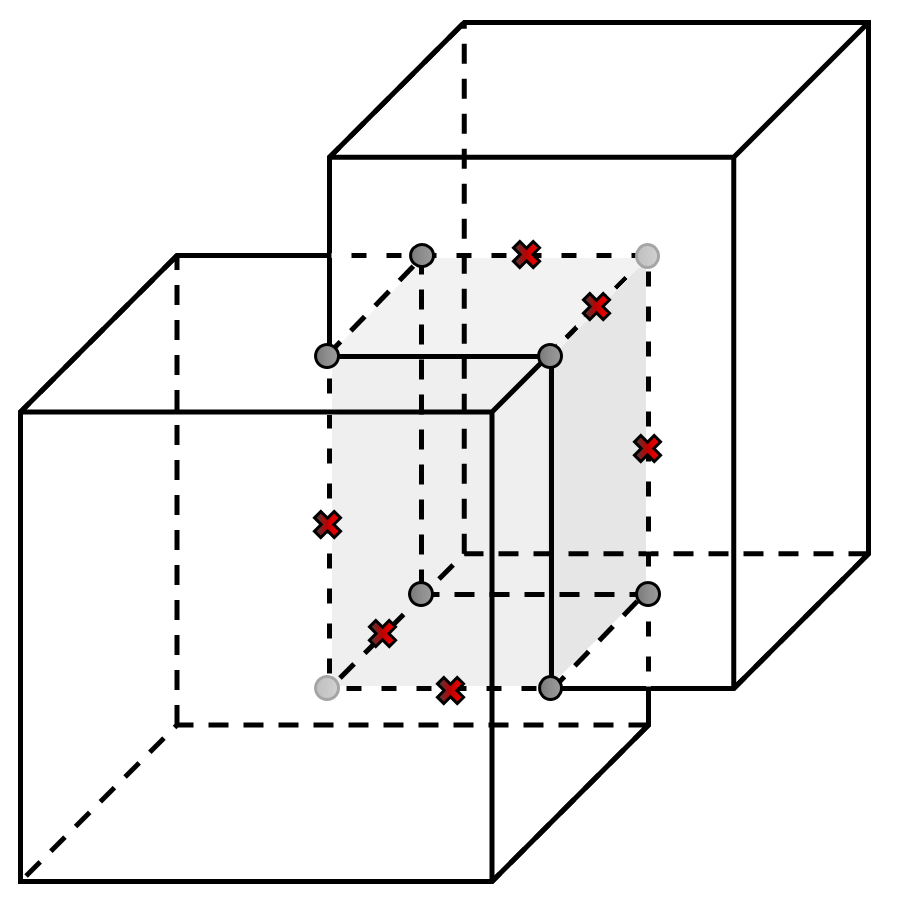} 
 	} 	
    \subfloat[P006][Resulting object\label{subfig:3D_BoolOP_OR_Rslt}]{
   		\includegraphics[scale=0.250]{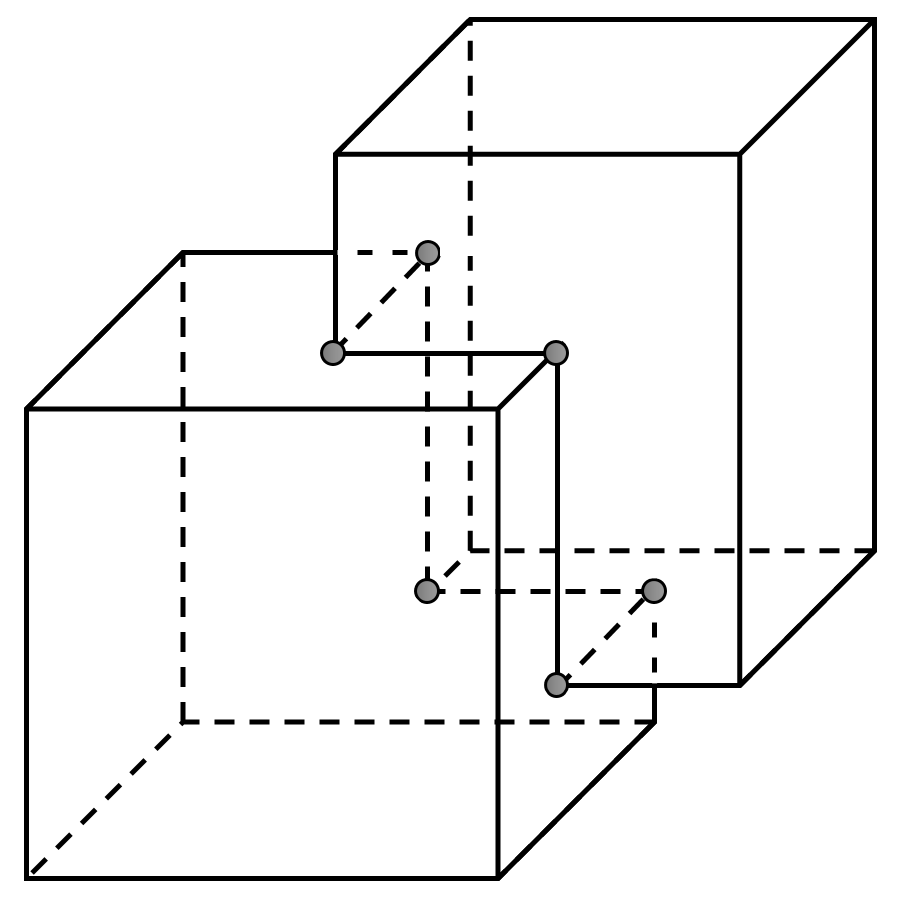} 
 	} 	
\caption{OR Boolean Operation (Union) in 3D\label{fig:figBOP3D}}
\end{figure}

The proposed algorithm can also be applied to a third dimension as shown in the example in Fig.~\ref{fig:figBOP3D}. In Fig.~\ref{subfig:3D_BoolOP_OR_Orgnl} the light gray region defines the overlap between both objects. Using one example of the operations for 2d structures, the application of the union (OR) strategy produces the result shown in Fig.~\ref{subfig:3D_BoolOP_OR_Rslt}.

\section{Conclusion}

This work presents an algorithm for the generation of full NFPs that is simple to implement, and deals with any degenerate cases without effort, such as perfect fits or perfect sliding placement positions. This algorithm also reduces problems with numerical precision that may arise in other approaches (such as orbital approaches, and other iterative processes). This method also presents the benefit of being easily extended for boolean operations in both 2D and 3D, with the same advantages as when generating the NFP. The development of algorithms as the one presented in this work are an important contribution to the field, by improving the robustness and quality of current approaches, and enabling new research paths that have been mainly unexplored due to their complexity (such as 3D irregular packing).

\renewcommand\refname{Bibliography}

\bibliographystyle{plainnat} 
\bibliography{myrefs} 

\begin{thebibliography}{18}
\providecommand{\natexlab}[1]{#1}
\providecommand{\url}[1]{\texttt{#1}}
\expandafter\ifx\csname urlstyle\endcsname\relax
  \providecommand{\doi}[1]{doi: #1}\else
  \providecommand{\doi}{doi: \begingroup \urlstyle{rm}\Url}\fi

\bibitem[Agarwal et~al.(2002)Agarwal, Flato, and Halperin]{kn:Agarwal2002}
P.K. Agarwal, E.~Flato, and D.~Halperin.
\newblock Polygon decomposition for efficient construction of minkowski sums.
\newblock \emph{Computational Geometry Theory and Applications}, 21:\penalty0
  29--61, 2002.

\bibitem[Babu and Babu(2001)]{kn:Babu2001}
A.R. Babu and N.R. Babu.
\newblock A generic approach for nesting of 2-d parts in 2-d sheets using
  genetic and heuristic algorithms.
\newblock \emph{Computer-Aided Design}, 33:\penalty0 879--891, 2001.

\bibitem[Bennell and Oliveira(2008)]{kn:Bennell2008a}
J.A. Bennell and J.F. Oliveira.
\newblock The geometry of nesting problems: A tutorial.
\newblock \emph{European Journal of Operational Research}, 184\penalty0
  (2):\penalty0 397--415, 2008.

\bibitem[Bennell and Oliveira(2009)]{kn:Bennell2009}
J.A. Bennell and J.F. Oliveira.
\newblock A tutorial in irregular shape packing problems.
\newblock \emph{Journal of the Operational Research Society}, 60:\penalty0
  93--105, 2009.

\bibitem[Bennell and Song(2008)]{kn:Bennell2008b}
J.A. Bennell and X.~Song.
\newblock A comprehensive and robust procedure for obtaining the no-fit-polygon
  using minkowski sums.
\newblock \emph{Computers \& Operations Research}, 35\penalty0 (1):\penalty0
  267--281, 2008.

\bibitem[Bennell et~al.(2001)Bennell, Dowsland, and Dowsland]{kn:Bennell2001}
J.A. Bennell, K.A. Dowsland, and W.B. Dowsland.
\newblock The irregular cutting-stock problem -- a new procedure for deriving
  the no-fit-polygon.
\newblock \emph{Computers and Operations Research}, 28:\penalty0 271--287,
  2001.

\bibitem[Burke et~al.(2006)Burke, Hellier, Kendall, and Whitwell]{kn:Burke2006}
E.~Burke, R.~Hellier, G.~Kendall, and G.~Whitwell.
\newblock A new bottom-left-fill heuristic algorithm for the two-dimensional
  irregular packing problem.
\newblock \emph{Operations Research}, 54:\penalty0 587--601, 2006.

\bibitem[Burke et~al.(2007)Burke, Hellier, Kendall, and Whitwell]{kn:Burke2007}
E.K. Burke, R.S.R. Hellier, G.~Kendall, and G.~Whitwell.
\newblock Complete and robust no-fit polygon generation for the irregular stock
  cutting problem.
\newblock \emph{European Journal of Operational Research}, 179\penalty0
  (1):\penalty0 27 -- 49, 2007.
\newblock ISSN 0377-2217.
\newblock \doi{http://dx.doi.org/10.1016/j.ejor.2006.03.011}.
\newblock URL
  \url{http://www.sciencedirect.com/science/article/pii/S0377221706001639}.

\bibitem[Cuninghame-Green(1989)]{kn:Cuninghame1989}
R.~Cuninghame-Green.
\newblock Geometry, shoemaking and the milk tray problem.
\newblock \emph{New Scientist}, 1677:\penalty0 50--53, 1989.

\bibitem[Dean et~al.(2006)Dean, Tu, and Raffensperger]{kn:Dean2006}
H.T. Dean, Y.~Tu, and J.F. Raffensperger.
\newblock An improved method for calculating the no-fit-polygon.
\newblock \emph{Computers \& Operations Research}, 33\penalty0 (6):\penalty0
  1521--1539, 2006.

\bibitem[Ghosh(1991)]{kn:Ghosh1991}
P.K. Ghosh.
\newblock An algebra of polygons through the notion of negative shapes.
\newblock \emph{CVGIP: Image Understanding}, 54\penalty0 (1):\penalty0
  119--144, 1991.

\bibitem[Gomes and Oliveira(2006)]{kn:Gomes2006}
A.M. Gomes and J.F. Oliveira.
\newblock Solving irregular strip packing problems by hybridising simulated
  annealing and linear programming.
\newblock \emph{European Journal of Operational Research}, 171\penalty0
  (3):\penalty0 811--829, 2006.

\bibitem[Li and Milenkovic(1995)]{kn:Li1995}
Z.~Li and V.~Milenkovic.
\newblock Compaction and separation algorithms for non-convex polygons and
  their applications.
\newblock \emph{European Journal of Operations Research}, 84:\penalty0
  539--561, 1995.

\bibitem[Mahadevan(1984)]{kn:Mahadevan1984}
A.~Mahadevan.
\newblock \emph{Optimization in computer aided pattern packing}.
\newblock PhD thesis, North Carolina State University, 1984.

\bibitem[Milenkovic et~al.(1991)Milenkovic, Daniels, and Li]{kn:Milenkovic1991}
V.J. Milenkovic, K.M. Daniels, and Z.~Li.
\newblock Automatic marker making.
\newblock In \emph{Proceedings of the Third Canadian Conference on
  Computational Geometry}, pages 243--246, 1991.

\bibitem[Sato et~al.(2013)Sato, Martins, and Tsuzuki]{kn:Tsuzuki2013}
A.K. Sato, T.C. Martins, and M.S.G. Tsuzuki.
\newblock Collision free region determination by modified polygonal boolean
  operations.
\newblock \emph{Computer-Aided Design}, 45\penalty0 (7):\penalty0 1029--1041,
  2013.

\bibitem[Watson and Tobias(1999)]{kn:Watson1999}
P.D. Watson and A.M. Tobias.
\newblock An efficient algorithm for the regular w1 packing of polygons in the
  infinite plane.
\newblock \emph{Journal of the Operational Research Society}, 50\penalty0
  (10):\penalty0 1054--1062, 1999.

\bibitem[Whitwell(2004)]{kn:Whitwell2004}
G.~Whitwell.
\newblock \emph{Novel heuristic and metaheuristic approaches to cutting and
  packing}.
\newblock PhD thesis, University of Nottingham, 2004.

\end{thebibliography}

\end{document}